\documentclass[12,preprint]{aastex}

\usepackage{amsmath}

 \newcommand{\bra}[1]{\left( #1 \right)}
 \newcommand{\sqb}[1]{\left[ #1 \right]}
 \newcommand{\super}[1]{$^{\text{#1}}$}
\newcommand{\sub}[1]{$_{\text{#1}}$}

\shorttitle{MonR2 - draft 13}
\shortauthors{M. Dierickx}

\begin{document}

\title{Submillimeter Array High-angular Resolution Observations of the Monoceros R2 Star Forming Cluster}
\author{M. Dierickx\altaffilmark{1}, I. Jim\'{e}nez-Serra\altaffilmark{3,2,1}, V. M. Rivilla\altaffilmark{4,5}, and Q. Zhang\altaffilmark{1}}
\email{{\it mdierickx@cfa.harvard.edu}}
\altaffiltext{1}{Harvard-Smithsonian Center for Astrophysics,
60 Garden St., Cambridge, MA 02138, USA; mdierickx@cfa.harvard.edu, qzhang@cfa.harvard.edu}
\altaffiltext{2}{European Southern Observatory, Karl-Schwarzschild-Str. 2, 85748, Garching, Germany; ijimenez@eso.org}
\altaffiltext{3}{Department of Physics and Astronomy, University College London, 132 Hampstead Road, London NW1 2PS, UK; i.jimenez@ucl.ac.uk}
\altaffiltext{4}{Centro de Astrobiolog\'{i}a (CSIC/INTA), Ctra. de Torrej\'{o}n a Ajalvir km 4, E-28850 Torrej\'{o}n de Ardoz, Madrid, Spain; ryvendel@gmail.com}
\altaffiltext{5}{Osservatorio Astrofisico di Arcetri, Largo Enrico Fermi, 5, I-50125, Firenze, Italia; rivilla@arcetri.astro.it}

\begin{abstract}
We present the first high-angular resolution study of the MonR2 star-forming complex carried out with the Submillimeter Array at (sub-)millimeter wavelengths. We image the continuum and molecular line emission toward the young stellar objects in MonR2 at 0.85~mm and 1.3~mm, with resolutions ranging from 0.5$"$ to $\sim 3$''. While free-free emission dominates the IRS1 and IRS2 continuum, dust thermal emission prevails for IRS3 and IRS5, giving envelope masses of $\sim$0.1-0.3 $M_\Sun$. IRS5 splits into at least two sub-arcsecond scale sources, IRS5B and the more massive IRS5A. Our \super{12}CO(2-1) images reveal 11 previously unknown molecular outflows in the MonR2 clump. Comparing these outflows with known IR sources in the IRS5 and IRS3 subclusters allows for tentative identification of driving stars. Line images of molecular species such as CH$_3$CN or CH$_3$OH show that, besides IRS3 (a well-known hot molecular core), IRS5 is also a chemically active source in the region. The gas excitation temperature derived from CH$_3$CN lines toward IRS5 is $144 \pm 15$~K, indicating a deeply embedded protostar at the hot-core evolutionary stage. SED fitting of IRS5 gives a mass of $\sim$7$\,$M$_\odot$ and a luminosity of 300$\,$L$_\odot$ for the central source. The derived physical properties of the CO outflows suggest that they contribute to the turbulent support of the MonR2 complex and to the gas velocity dispersion in the clump's center. The detection of a large number of CO outflows widespread across the region supports the competitive accretion scenario as origin of the MonR2 star cluster.
\end{abstract} 

\section{Introduction}
\label{sec:intro}

Massive young stars are found almost exclusively in clusters with stars of lower stellar masses. How molecular clouds collapse and fragment to assemble stars of distributed masses has been the subject of intensive studies in the past several years. There are currently two main theories of massive star and star cluster formation. While the first scenario involves the monolithic gravitational collapse of a turbulent-supported massive core at high accretion rates \citep[e.g.][]{yorke02,mckee03,krumholz09}, the second scenario - the competitive accretion scenario - initially forms a cluster of low- and intermediate-mass stars where the most massive stars grow via Bondi-Hoyle accretion at the center of the cluster potential well \citep[see e.g.][]{bonnell06}. In the competitive accretion scenario, it is thus expected that massive stars form at the center of the cluster with the stellar density of low-mass stars peaking at this position. At the cluster center where stellar densities are predicted to be high, dynamical encounters may take place as proposed for the Orion BN/KL and the DR21 high-mass star forming regions \citep[see e.g.][]{zapata09,zapata13,rivilla13a,rivilla14}. Therefore, to study the formation of massive stars, it is crucial to better understand the distribution of low-mass stars and their impact on the subsequent evolution of high-mass star forming clusters.

Molecular outflows appear early in the star formation process as a mechanism to remove the excess angular momentum in the star/circumstellar disk system produced by gas accretion. Molecular outflows thus represent a useful tool to probe the population of Class 0 low-mass stars in clusters. Since molecular outflows inject momentum and energy into the surrounding medium, they are expected to increase the level of turbulence and even disrupt the parental molecular cloud, potentially impacting any subsequent star formation within the cluster \citep[see][]{stanke07,rivilla13b}.

The Monoceros R2 (MonR2) molecular cloud complex is located at a distance of 830 pc \citep{herbst76} and extends over 3$^\circ$$\times$6$^\circ$ in the plane of the sky. MonR2 contains several sites of star formation \citep[see][and references therein]{carpenterreview08}. The MonR2 clump (size of $\sim$3$'$ or 0.7$\,$pc) has a mass of 1800$\,$M$_\odot$ \citep{ridge03} and is the most active star forming region in the MonR2 complex as revealed by the blister-type HII region \citep{massi85,wood89}, the dense stellar cluster \citep{carpenter97,kohno02}, the large-scale CO outflow \citep[one of the largest - $\sim$0.6-7 pc - and most massive - 275$\,$M$_\odot$ - reported so far;][]{loren81,wolf90,tafalla97}, and the H$_2$O and OH masers found in this region \citep{downes75,genzel77,smits98}. Among the IR sources detected in the cluster \citep{carpenter97}, the sources IRS1, IRS2 and IRS3 are the most luminous objects in the MonR2 region, with luminosities of 3000$\,$L$_\odot$, 6000$\,$L$_\odot$ and 15000$\,$L$_\odot$ respectively \citep{henning92}. The ionizing source of the HII region is found 4$"$ away from the peak of the radio continuum emission \citep{aspin90} and it is resolved into two components, IRS1 NE and IRS1 SW. While IRS1 NE is a foreground field star \citep{howard94}, IRS1 SW is responsible for the HII region and it is likely a B0-type star in the ZAMS \citep{massi85}. IRS2 appears as a bright source in the K-band, indicating that this object is embedded in a molecular cloud \citep{beckwith76,aspin90}. IRS2 is very compact and it does not show any structure at sub-arcsecond scales \citep{alvarez04,jimenez-serra13}. This source is responsible for the reflection nebula seen in the K-band and $nbL$ images \citep{carpenter97} and coincident with the edges of the radio continuum emission mapped with the VLA \citep{massi85}. The IRS3 source is the most luminous object in the region and, as revealed by Speckle interferometry, it is in fact a cluster of IR sources, one of which (IRS3B) shows a microjet \citep{preibisch02}. Since the orientation of this jet (position angle of P.A.$\sim$50$^\circ$ from north to east) is roughly perpendicular to the large-scale CO outflow, it is unlikely that this source (IRS3B) is the driving object of the outflow. In fact, it still remains unknown which source within the MonR2 cluster is the actual driving source of the 7$\,$pc-long CO outflow in MonR2 \citep{giannakopoulou97,tafalla97}. 

Other less luminous IR sources are also detected toward the MonR2 cluster, for example IRS4-7, a$_{\rm N}$, a$_{\rm S}$, and b-h \citep[see ][]{beckwith76,hackwell82,aspin90, carpenter97}. Physical properties have only been constrained for a few of them: total luminosity estimates of  $\sim$700-800~L$_\odot$ and $\sim$300~L$_\odot$ exist for IRS4 and IRS5, respectively \citep{henning92,hackwell82}. More extensive studies, especially at high-angular resolution and across several wavelengths, are needed to fully characterize the properties of these intermediate-luminosity sources associated with the MonR2 cluster.

By using near-IR 2.12$\,$$\mu$m H$_2$ observations, \citet{hodapp07} reported the detection of 15 new H$_2$ jets in the MonR2 cluster that are likely associated with young Class 0 and I sources at their main accretion phase. These jets were mainly found toward the edges of the MonR2 molecular cloud, which was interpreted as a signature of triggered star formation \citep{hodapp07,carpenterreview08}. However, near-IR observations may provide a biased view of the global population of molecular outflows in deeply embedded star forming clusters such as MonR2, because of the large extinction found in these regions. Alternatively, the rotational transitions of carbon monoxide (CO) in the millimeter and sub-millimeter wavelength range are known to be excellent probes of the material swept-up by the propagation of the jet into the surrounding molecular envelope \citep[e.g.][]{gueth99,lee07}. Interferometric observations of the rotational transitions of CO at sub-millimeter/millimeter wavelengths are thus needed to unveil the most embedded (and the youngest) population of molecular outflows, and therefore of Class 0 and Class I low-mass stars, at the cluster centers of high-mass star forming regions.  

In this paper, we report the first interferometric images of the thermal continuum and molecular line emission carried out with the Submillimeter Array \citep[SMA;][]{ho04}\footnote{The Submillimeter Array is a joint project between the Smithsonian Astrophysical Observatory and the Academia Sinica Institute of Astronomy and Astrophysics and is funded by the Smithsonian Institution and the Academia Sinica.} toward the MonR2 high-mass star forming cluster at angular resolutions ranging from 0.5$"$ to $\sim$3$"$ (i.e. from 0.002 to 0.012$\,$pc at a distance of 830$\,$pc)\footnote{MonR2 has only been imaged once before in HCO+ $J$=1$\rightarrow$0 by using the Hat Creek interferometer and at an angular resolution of $\sim$9'' \citep{gonatas12}.}. Our SMA images of the $^{12}$CO $J$=2$\rightarrow$1 line emission reveal the presence of 11 new CO outflows toward the innermost $\sim$1$'$-region in the cluster. This population of outflows is widely distributed across the MonR2 clump, which contradicts the proposed scenario of triggering for the MonR2 star cluster. The paper is organized as follows. In Section 2, we report the details about the SMA observations. The results of the dust continuum and $^{12}$CO line emission are presented in Sections 3.1 and 3.2, respectively. In Section 3.3 we analyze the line transitions of other molecular tracers such as CH$_3$CN, CH$_3$OH or SO$_2$ detected toward the IRS3 and IRS5 sources, and in Section 4 we discuss our results. The conclusions are summarized in Section \ref{sec:conclusions}.

\section{Observations and data reduction}
\label{sec:observations}

\begin{deluxetable}{cccccccccc}
\tabcolsep 3pt 
\tabletypesize{\scriptsize}
\tablewidth{0pt}
\tablecaption{Instrumental parameters of the SMA observations.}
\tablehead{
\colhead{Date} & \colhead{Config.} & \colhead{L.O. freq. (GHz)} & \colhead{Synthesized Beam} & 
\colhead{$\tau_{225\text{GHz}}$} & \colhead{$T_{sys}$ (K)} & \colhead{BP Cal.} & \colhead{Flux Cal.} & \colhead{Gain Cal.}
}
\startdata
Feb 13, 2010 & VEX & 224.611 & $0.68''\times0.45''$, P.A.=-$43^\circ$ & 0.05 & 200-240 & 3C273 & Titan & 0607-085/0730-116 \\
Feb 20, 2010 & VEX & 224.611 & $0.68''\times0.45''$, P.A.=-$43^\circ$  & 0.03 & 200-240 & 3C273 & Titan & 0607-085/0730-116 \\
Nov 15, 2011 & Compact & 226.227 & $2.75''\times2.73''$, P.A.=-$77^\circ$ & 0.07 & 200-240 & BLLAC & Ganymede & 0607-085/0530+135 \\
Nov 15, 2011 & Compact & 348.324 & $2.29''\times1.50''$, P.A.=-$39^\circ$ & 0.07 & 400-500 & BLLAC & Ganymede & 0607-085/0530+135 \\
\enddata
\label{tab:observations}
\end{deluxetable}

Observations of the MonR2 star forming region were carried out with the SMA at 1.3~mm in very extended configuration (VEX) in February 2010 for two tracks, and in Compact configuration in November 2011 for one track. Because the Compact track was performed in dual receiver mode, we simultaneously imaged MonR2 at 0.85~mm. The observations were phase-centered on IRS2 at coordinates $\alpha$(J2000) = $06^{\text{h}}07^{\text{m}}45.83^{\text{s}}$ and $\delta$(J2000) = $-06^{\circ}22'53.5''$ \citep{carpenter97}, and the central radial velocity of the source was set at 10 km~s$^{-1}$ \citep{torrelles83}. Instrumental and calibration parameters for the VEX and Compact observations are summarized in Table \ref{tab:observations}. The VEX observations were obtained in single-receiver mode with a total bandwidth of 4~GHz per sideband. The resulting frequency coverage was 216.64-220.60~GHz and 228.60-232.58~GHz. For the Compact observations, the receivers were configured in dual receiver mode with a total bandwidth of 2~GHz per sideband. As a result the frequency coverage was in 2~GHz windows, as follows: 220.24-222.23~GHz and 230.24-232.23~GHz for the 230~GHz receiver, and 342.67-344.66~GHz and 352.67-354.66~GHz for the 400~GHz receiver. The high-frequency receiver on one antenna was malfunctioning during the Compact track and the resulting data flagged, causing the beam to be elongated at 400~GHz. The correlator provided a uniform channel spacing of 0.8~MHz, yielding a channel width of $\sim1.1$~km~s$^{-1}$ at 1.3~mm (230~GHz) and $\sim0.7$~km~s$^{-1}$ at 0.85~mm (400~GHz). The primary beam (field of view) of the observations was $\sim$57'' at 230~GHz and $\sim$35'' at 400~GHz. Calibration of the raw data was done with the IDL MIR software package, while continuum subtraction, imaging and deconvolution were carried out with MIRIAD. We note that for the Compact data, the quasar BLLAC was used as bandpass calibrator (see Table \ref{tab:observations}). Since BLLAC's spectrum shows a prominent \super{12}CO absorption feature, the bandpass calibration of the corresponding upper sideband was performed by phase-conjugating the lower sideband SMA BLLAC data.

\section{Results}
\label{sec:results}

\subsection{Dust Continuum Emission}
\label{subsec:Dustcontinuum} 

\subsubsection{Compact SMA Observations}
\label{subsubsec:COMcontinuum}

\begin{deluxetable}{cccccccc}
\tabletypesize{\footnotesize}
\tablewidth{0pt}
\tablecaption{Properties of the dust continuum emission in Compact configuration.}
\tablehead{
\colhead{Source} & \colhead{Wavelength} & \colhead{$\alpha$ (J2000)} & \colhead{$\delta$ (J2000)} & \colhead{Size\tablenotemark{a} (''$\times$'')} & \colhead{P.A. ($^{\circ}$)} & \colhead{$F_{peak}$\tablenotemark{b} (Jy/beam)} & \colhead{$F_{int}$\tablenotemark{b} (Jy)}
}
\startdata
IRS2 & 1.3 mm & $06^\text{h}07^\text{m}45.803^\text{s}$ & $-06^{\circ}22'53.51''$ & $3.15\times2.78$ & 75 &  $0.154\pm0.005$  & $0.180\pm0.005$ \\
IRS2 & 0.85 mm & $06^\text{h}07^\text{m}45.800^\text{s}$ & $-06^{\circ}22'53.55''$ & $3.00\times2.79$ & 140 & $0.139\pm0.004$ & $0.154\pm0.005$ \\
\hline
IRS5 & 1.3 mm & $06^\text{h}07^\text{m}45.588^\text{s}$ & $-06^{\circ}22'39.42''$ & $3.14\times2.79$ & 122 & $0.109\pm0.005$ & $0.127\pm0.006$ \\
IRS5 & 0.85 mm & $06^\text{h}07^\text{m}45.612^\text{s}$ & $-06^{\circ}22'39.47''$ & $2.95\times2.86$ & 10 & $0.389\pm0.012$ & $0.436\pm0.013$ \\
\hline
IRS3 & 1.3 mm & $06^\text{h}07^\text{m}47.811^\text{s}$ & $-06^{\circ}22'56.18''$ & $3.37\times2.88$ & 55 & $0.059\pm0.008$ & $0.077\pm0.011$ \\

\enddata
\tablenotetext{a}{These are non-deconvolved source sizes.}
\tablenotetext{b}{The errors in the peak flux and total density flux correspond to those obtained from the two-dimensional Gaussian fits of the emission. The 0.85~mm image was spatially smoothed to the same angular resolution as that of the 1.3~mm image.}
\label{tab:COMcontinuum}
\end{deluxetable}

\begin{figure*}
\begin{center}
\includegraphics[scale=0.6, angle=0]{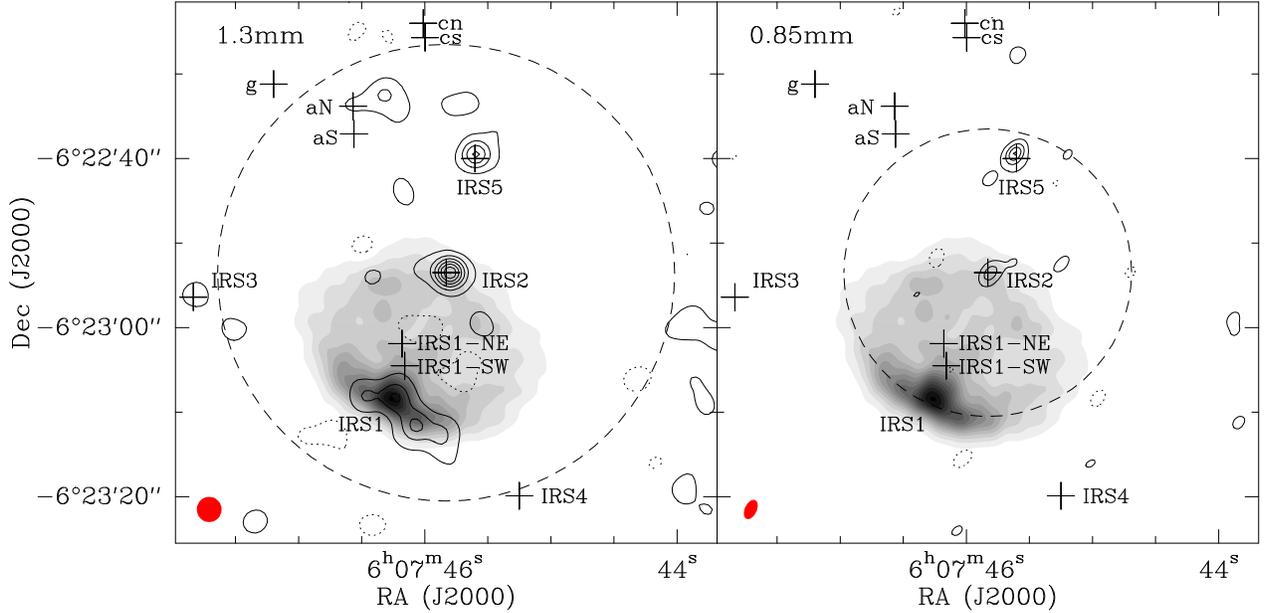} 
\caption{Left panel: Continuum emission detected at 1.3~mm in Compact configuration toward the MonR2 star forming region (contours) overlaid on the 1.3~cm continuum image \citep[from the dataset of][]{zapata09}. Contours start at 12~mJy beam\super{-1} ($3\sigma$) and are spaced at 24~mJy beam\super{-1} ($6\sigma$) intervals. Dotted contours correspond to the negative $3\sigma$ level. The beam size is shown as a red ellipse in the lower left corner. For the VLA data, the first contour is at 3~mJy beam\super{-1} ($3\sigma$) and the step level is 17~mJy beam\super{-1} ($17\sigma$). The dashed circle indicates the primary beam of the observations, $\sim$57'' at 1.3~mm. Right panel: As for the left panel, but for the 0.85~mm continuum image. The first contour is at 30~mJy beam\super{-1} ($3\sigma$) and the step level is 60~mJy beam\super{-1} ($6\sigma$). At 0.85~mm the primary beam of the observations is $\sim$35'' (dashed circle). The plus signs indicate the positions of the main sources in the region, including IRS2, IRS3, IRS4, IRS5 and the ionizing star of the HII region, IRS1 SW \citep[note that this is different from IRS 1 NE, which is an unrelated non-embedded star detected in the optical; see][]{cohen77}. The maps were made using uniform weighting.
\label{fig:continuum} }
\end{center}
\end{figure*}

\begin{deluxetable}{ccc}
\tablewidth{0pt}
\tablecaption{Envelope mass estimates from COM data. }
\tablehead{Source & \multicolumn{2}{c}{Mass ($M_{\Sun}$)} \\ \cline{2-3} 
& T\sub{dust} = 50 K & T\sub{dust} = 150 K}
\startdata
IRS5 & 0.26 & 0.08 \\
IRS3 & 0.15 & 0.05 \\
\enddata
\tablecomments{
For IRS3, no emission is detected at 0.85~mm, so the mass is derived using the same dust opacity index as for IRS5.}
\label{tab:commass}
\end{deluxetable}

In Figure \ref{fig:continuum}, we present the continuum emission seen in Compact configuration at 1.3~mm (left panel) and at 0.85~mm (right panel) towards MonR2. For comparison, Figure \ref{fig:continuum} includes the 1.3~cm Very Large Array continuum image from the dataset of \citet{zapata09}, where the MonR2 central HII region is clearly visible. Our SMA image at 1.3~mm shows extended structure toward IRS1 coincident with the brightest emission seen at centimeter wavelengths. However, in the 0.85~mm map of Figure 1 this source is not detected. This could be due to: i) IRS1 is located at the edge of the primary beam of the 0.85~mm observations; and ii) the continuum emission detected in the SMA images probes the innermost regions around IRS1 and therefore arises from ionized gas associated with the UC HII region. Although subject to large uncertainties, the comparison between the 1.3 cm and 1.3 mm flux for IRS1 yields a negative spectral index $\alpha$$\sim$$-$0.5 (with $S_{\nu}$$\propto\nu^{\alpha}$), consistent with free-free emission. By using this spectral index, we can extrapolate the expected flux for this source to 0.85 mm. This flux, at the half-power of the primary beam, is $\sim$20$\,$mJy$\,$beam$^{-1}$, well below the 3$\sigma$ noise level of 30 mJy$\,$beam$^{-1}$ in the 0.85 mm image of Figure$\,$\ref{fig:continuum}. This explains why the IRS1 source has not been detected in our SMA continuum image at 0.85 mm.

From Figure 1, we find that the brightest sources in our SMA images are IRS2 and IRS5. A qualitative comparison of the two maps at 1.3~mm and 0.85~mm reveals that these two sources are very different in nature. While IRS2 appears brighter at 1.3~mm than at 0.85~mm (as for IRS1 and as expected for an HII region), IRS5 shows the opposite behavior, indicating that its emission is mainly due to dust. In Table \ref{tab:COMcontinuum} we report parameters calculated from two-dimensional Gaussian fits to the continuum emission of these two sources. The derived coordinates are close to those obtained from previous infrared and X-ray measurements \citep[e.g.][]{carpenter97, kohno02, nakajima03}. As IRS5 is offset from the phase center of the observations by approximately 15'', primary-beam corrected maps were used to estimate its flux parameters. After smoothing the 0.85~mm images to the same angular resolution as that of the 1.3~mm data ($\sim3$''), we compare the peak flux values at 1.3~mm and 0.85~mm (see Table \ref{tab:COMcontinuum}) to estimate the spectral index, $\alpha$, of the two sources (with $S_\nu \propto \nu^\alpha$). We obtain $\alpha_{\text{IRS2}}  \sim -0.2$ and $\alpha_{\text{IRS5}} \sim 3.0$. The decreasing spectral index of IRS2 is consistent with that previously reported by \citet{jimenez-serra13}, and indicates that this source is dominated by optically thin free-free emission.  This source, with an estimated central mass of 13~M\sub{$\Sun$} (a B1-type star in the ZAMS) and a luminosity of 6000~L\sub{$\Sun$} \citep{henning92}, likely powers a dense and collimated ionized jet \citep{jimenez-serra13}.

In contrast to IRS2, the spectral index of IRS5 steeply increases with frequency ($\alpha_{IRS5}=3.0$), indicating that its emission at millimeter wavelengths is dominated by dust thermal emission. This suggests that IRS5 is at an earlier stage of evolution than IRS2. Since the emission from IRS5 is dominated by dust, we can estimate the envelope dust mass in the source following the method from \citet{hildebrand83}:
\begin{equation}
M_{\text{dust}} = \frac{F(\nu) d^2}{B_{\nu} (T_{\text{dust}}) \kappa_\nu}
\end{equation}
\label{eq:dustmass}
where $F(\nu)$ is the continuum flux density at frequency $\nu$, $d$ is the distance to the source, $B_{\nu}(T_{\text{dust}})$ is  the Planck function for dust temperature $T_{\text{dust}}$, and $\kappa_\nu$ is the dust opacity. $\kappa_\nu$ (cm$^2$ g$^{-1}$) is given by $10 \bra{\frac{\nu}{1.2\text{THz}}}^{\beta}$ , where $\beta$ is the dust opacity index. Assuming a power law of the type $Q(\nu) \propto \nu^{\beta}$ for the emissivity $Q(\nu)$,  in the Rayleigh-Jeans limit $\beta$ and $\alpha$ are then related by $\beta = \alpha - 2$. Since $\alpha_{\text{IRS5}} \sim 3.0$, we therefore obtain $\beta$\sub{IRS5}~$\sim 1$. Although this value is similar to those assumed by \citet{qiu07} for a sample of massive star-forming regions, we cannot rule out the possibility that our estimate of $\beta$ for IRS5 is affected by a small contribution from free-free emission, especially at 1.3~mm \citep[see][]{giannakopoulou97}. As an upper limit for the dust temperature, we use the gas temperature derived toward IRS5 from the rotational diagram of the CH\sub{3}CN (12-11) K = 0 to K = 6 lines detected with the SMA ($T\sim150$~K; see Section \ref{subsec:molecular}). As a lower limit for the dust temperature, we assume $T$\sub{dust} = 50~K, consistent with the $T$\sub{dust} values derived by \citet{thronson80} towards IRS5 and with those estimated by \citet{sridharan02} towards high-mass star forming regions such as IRAS18264-1152 and IRAS23151+5912. 

From the integrated flux reported in Table \ref{tab:COMcontinuum}, and considering a constant gas-to-dust mass ratio of 100 \citep[following e.g.][]{giannakopoulou97,beuther07}, we obtain estimates for the IRS5 envelope gas mass of $\sim0.08-0.3$~M\sub{$\Sun$}, as reported in Table \ref{tab:commass}. We note that these masses are smaller than those reported at poorer angular resolution (16''-27'') for IRS5 by \citet[][about 5~M\sub{$\Sun$}, see clump C5 in their Table 4]{giannakopoulou97}. This is due to the fact that the SMA only probes the densest regions toward IRS5, filtering out most of the extended emission around this source. In Table~$\,$\ref{tab:COMcontinuum}, we also report the continuum flux measured at 1.3 mm toward IRS3 (after the primary-beam correction of the images). Since this source falls outside the primary beam of our observations, the measured 1.3 mm flux and derived dust continuum properties (Table~$\,$\ref{tab:commass}) should be taken with caution.

Toward the north-east of IRS5 we also find some continuum emission in the SMA 1.3 mm continuum image at the 9$\sigma$ level (Figure$\,$\ref{fig:continuum}). Although the bulk of the emission does not coincide with the location of any IR source, the elongated asymmetry toward the east of this emission is indeed associated with source a$_{\rm N}$ detected by \citet{carpenter97} and \citet{andersen06}. As shown below, this source is resolved into several condensations (see Figure$\,$\ref{fig:VEXCom}) and could therefore represent a cluster of deeply embedded protostars. Future observations covering this region will help to establish the origin of the 1.3 mm continuum emission toward the northeast of IRS5.

\subsubsection{VEX Observations of the Small-Scale Structure}
\label{subsubsec:VEXcontinuum}

\begin{deluxetable}{lcc}
\tablewidth{0pt}
\tablecaption{IRS5 properties with combined Compact and VEX data at 1.3~mm. }
\tablehead{ & IRS5A & IRS5B}
\startdata
R.A. (J2000) & $06^\text{h}07^\text{m}45.60^\text{s}$ & $06^\text{h}07^\text{m}45.65^\text{s}$ \\
DEC (J2000) & $-06^{\circ}22'39.37''$ & $-06^{\circ}22'40.07''$ \\
Source size (''$\times$'', PA) & 0.90$\times$0.70, 44$^{\circ}$ & $\leq$0.7$"$ \\
$F_{\rm peak}$ (Jy/beam) & $0.080\pm0.002$ & $0.015\pm0.003$ \\
$F_{\rm int}$ (Jy) & $0.124\pm0.003$ & $\ldots$ \\
Mass (T\sub{dust} = 150 K) & 0.07 $M_{\Sun}$&  0.01 $M_{\Sun}$\\
Mass (T\sub{dust} = 50 K) & 0.24 $M_{\Sun}$& 0.03 $M_{\Sun}$ \\
\enddata
\tablecomments{
The errors in the peak flux and total density flux correspond to those obtained from the Gaussian fits of the emission. For IRS5B the peak intensity was used to derive a mass estimate because the source is point-like and no reliable Gaussian fit was obtained. The source sizes are not beam-deconvolved. }
\label{tab:COMVEXcontinuum}
\end{deluxetable}

\begin{figure*}
\begin{center}
\includegraphics[scale=0.6]{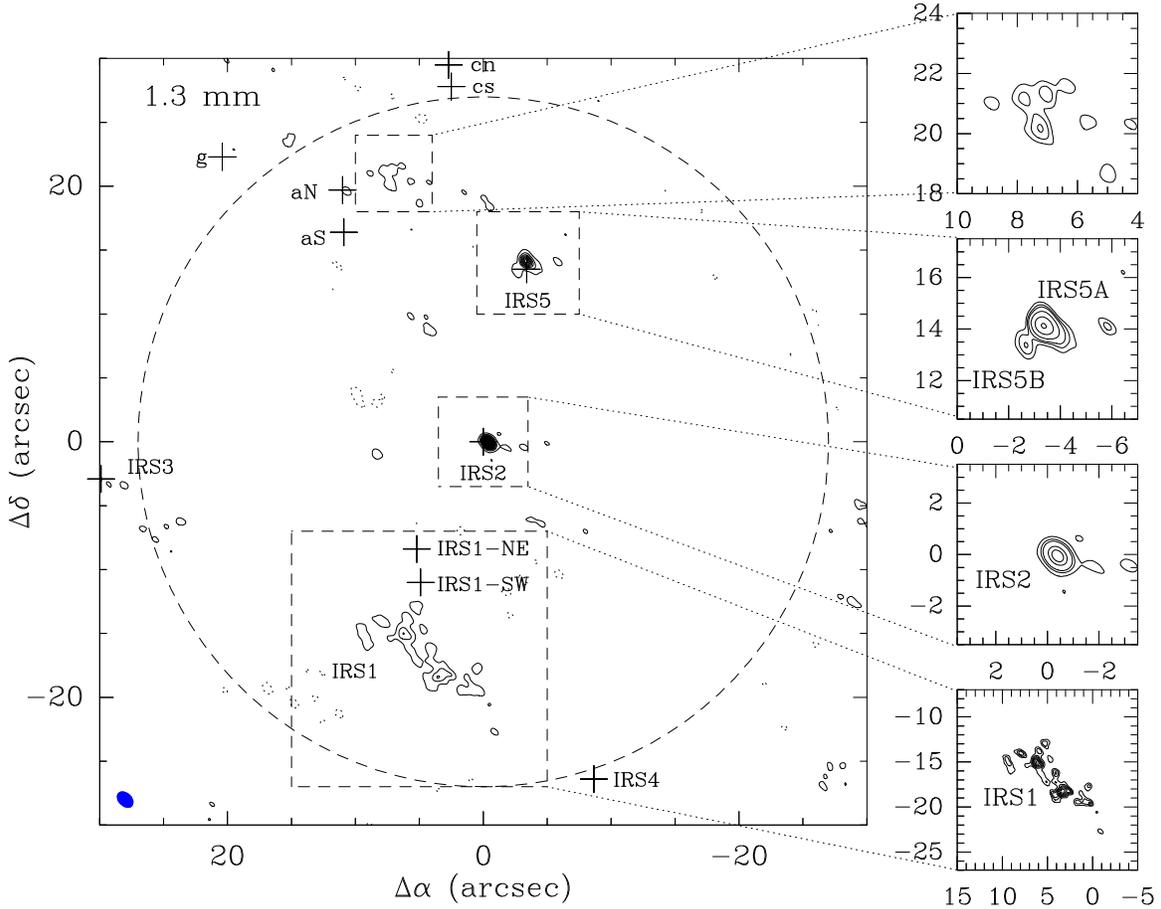} 
\caption{Combined VEX and Compact data at 1.3~mm. Left panel: In this map of the entire field of view, contours start at $3\sigma$ and are spaced at $3\sigma$ intervals, with $\sigma \sim 2.5$~mJy/beam. The beam size of the map is shown as a blue ellipse in the bottom left corner of the image. Large crosses indicate the position of the main sources in the region as in Figure \ref{fig:continuum}. For the top two zoom-in panels, we show the $3\sigma$, $4\sigma$, $5\sigma$, $8\sigma$, $15\sigma$, and $25\sigma$ contours. For IRS2, we show the $3\sigma$, $6\sigma$, $10\sigma$, $30\sigma$, $50\sigma$, and $70\sigma$ contours. Finally, for IRS1 we show contours starting at $3\sigma$ and spaced by $1\sigma$. Dotted contours correspond to the negative $3\sigma$ level. The map was made using uniform weighting.
\label{fig:VEXCom} }
\end{center}
\end{figure*}

\begin{figure*}
\begin{center}
\includegraphics[scale=0.6]{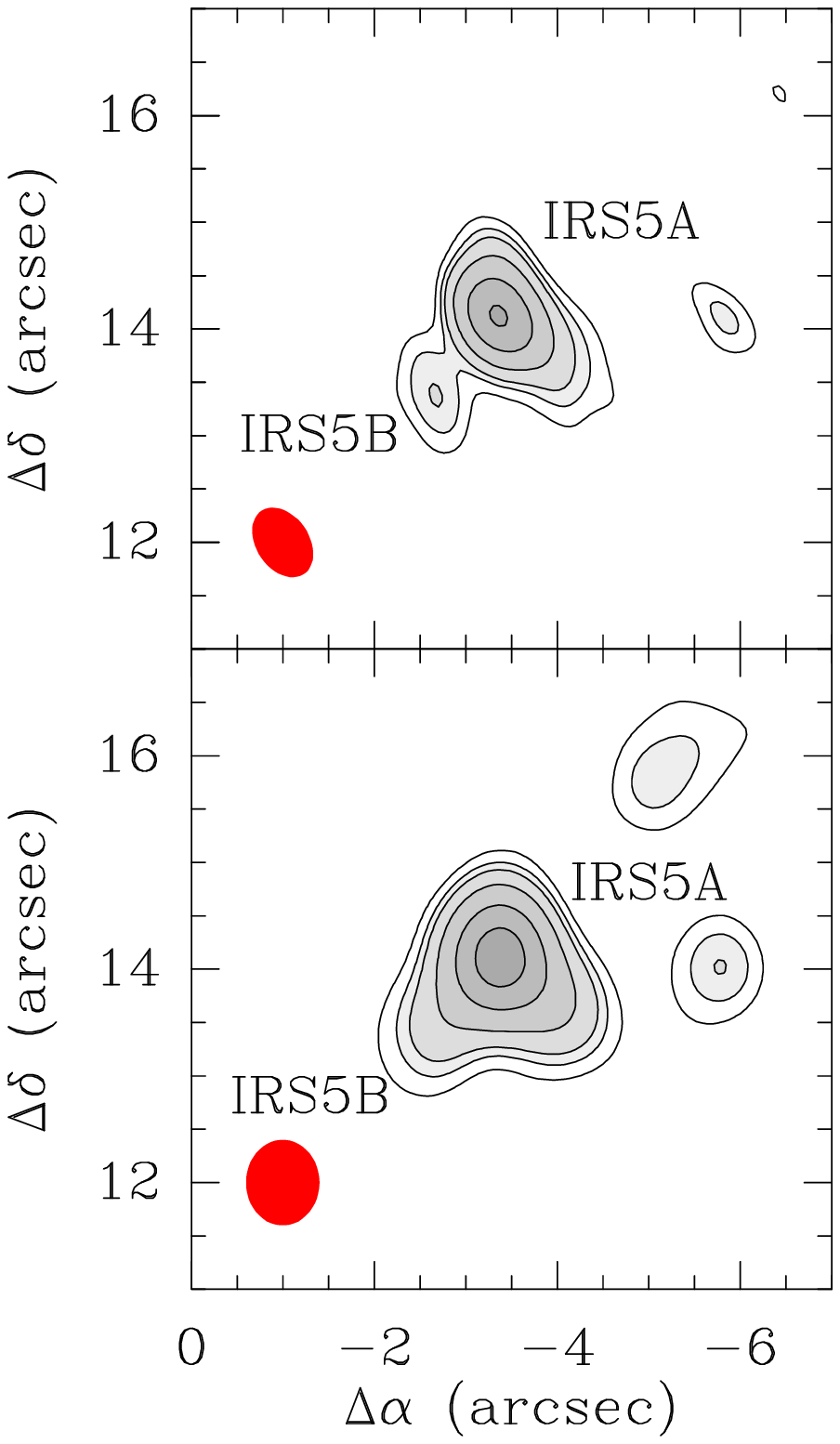} 
\caption{Maps of the combined VEX and Compact 1.3~mm data for IRS5 produced with different beams. The beam size of the map is shown as a red ellipse in the bottom left corner of each panel. The elongation along the northeast-southwest axis of IRS5 remains when the image is convolved with a circular beam. We show the $3\sigma$, $4\sigma$, $5\sigma$, $8\sigma$, $15\sigma$, and $25\sigma$ contours, with $\sigma \sim 2.5$~mJy/beam. The map was made using uniform weighting.
\label{fig:IRS5circbeam} }
\end{center}
\end{figure*}

Figure \ref{fig:VEXCom} presents the 1.3~mm continuum map of the MonR2 cluster obtained by combining the SMA data at 1.3~mm in the Compact and VEX configurations. The VEX data provide a much improved angular resolution of 0.5''. Toward IRS1, the extended emission associated with this infrared source shows a complex structure with two bright emission peaks with peak intensities detected at the 9$\sigma$ level. These peaks, however, do not coincide with the IR sources IRS1 NE or IRS1 SW previously reported by \citet{aspin90} and \citet{howard94} and shown by black crosses in Figure \ref{fig:VEXCom}; rather, they appear associated with the IR ring nebula related to the HII region \citep{carpenter97,andersen06,dewit09}. This indicates that the 1.3~mm emission is likely associated with dense inhomogeneities in the ionized gas. 

Figure \ref{fig:VEXCom} also shows that IRS2 remains unresolved by the present 0.76" $\times$ 0.54" beam. The secondary continuum peak toward the southwest of IRS2 is not associated with a real source, but rather due to limited UV sampling. This implies that the source does not include substructure at scales of 0.5$"$ (i.e. 400~AU at a distance of 830~pc). This compact morphology is consistent with the results of \citet{alvarez04} obtained by using near-IR Speckle imaging.

Further north, IRS5 shows hints of substructure not seen previously in the Compact data alone. In particular, a secondary peak to the southeast of the main source appears as a $5\sigma$ detection. For the purpose of characterizing these sources, we refer to the main IRS5 peak as IRS5A, and to the secondary source as IRS5B. Flux parameters for these two sources are reported in Table \ref{tab:COMVEXcontinuum}. Following the method in \citet{hildebrand83} and assuming a dust opacity index of $\beta\sim1$, we estimate envelope gas masses of 0.07-0.24~M\sub{$\Sun$} for IRS5A and of 0.01-0.03~M\sub{$\Sun$} for IRS5B (see Table \ref{tab:COMVEXcontinuum}.) As expected, IRS5A is the dominant source, as it accounts for $\sim90$\% of the total gaseous envelope mass in IRS5 measured in Compact configuration. In addition, an elongation along the northeast-southwest axis is highlighted by the contours, suggesting the existence of a third source that remains unresolved at the current resolution. As shown in Figure~\ref{fig:IRS5circbeam}, this elongation is robust when the image is convolved with a circular beam. Finally, the continuum peak to the west of IRS5A might be another source. However, since the emission is only detected at the 4 sigma level, higher sensitivity observations are required for confirmation (Figure~\ref{fig:IRS5circbeam}, upper panel).

\subsection{$^{12}$CO Emission: Outflowing Gas in MonR2}
\label{subsec:12co} 

\begin{figure*}
\begin{center}
\includegraphics[scale=0.6]{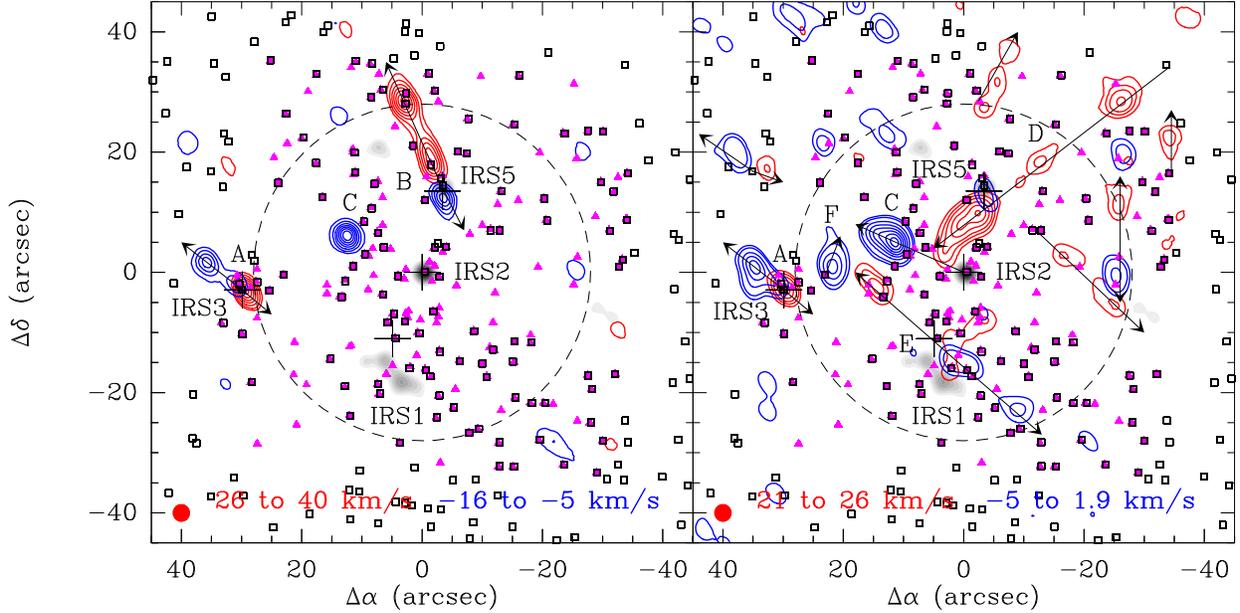} 
\caption{Moment-zero maps of the CO(2-1) transition in MonR2 (colored contours) overlaid on the 1.3~mm continuum (greyscale). Left panel: The redshifted emission has been integrated over the 26-40~km~s$^{-1}$ velocity range ($\sigma \sim 0.15$~Jy/beam km~s$^{-1}$); while the blueshifted emission has been integrated over the velocity range  -16 to -5~km~s$^{-1}$ ($\sigma \sim 0.14$~Jy/beam km~s$^{-1}$).  The first contour and step levels are $4\sigma$. Right panel: The redshifted gas appears for velocities comprised between 21 and 26~km~s$^{-1}$ ($\sigma \sim 0.60$~Jy/beam km~s$^{-1}$). The 3$\sigma$, 5$\sigma$ and 7$\sigma$ contours are shown; contour spacing is $3\sigma$ thereafter. The blueshifted counterparts appear between -5~km~s$^{-1}$ and 1.9~km~s$^{-1}$ ($\sigma \sim 0.41$~Jy/beam km~s$^{-1}$). Here the 3$\sigma$, 5$\sigma$ and 10$\sigma$ contours are shown; contour spacing is $5\sigma$ thereafter. The brightest outflows are labeled with letters. Magenta triangles indicate the positions of the IR sources observed with NICMOS by \citet{andersen06}, while black squares mark the coordinates of the IR sources identified by \citet{carpenter97}. Black arrows indicate the conjectured orientation of the outflows. The beam size is shown as a red ellipse in the lower left corner, and the dashed circle indicates the primary beam of the observations, $\sim$57'' at 1.3~mm. The maps were made using uniform weighting.
\label{fig:12cooutflows} }
\end{center}
\end{figure*}

\subsubsection{Morphology}
\label{subsubsec:12comorphology}

Figure \ref{fig:12cooutflows} presents the integrated intensity maps of the \super{12}CO(2-1) emission (at 230.538~GHz) measured toward MonR2 in compact configuration and overlaid on the continuum emission at 1.3~mm. The \super{13}CO(2-1) line transition was also observed within our frequency setup and it will be used in Section~\ref{subsubsec:12coproperties} to determine the optical depth of the \super{12}CO outflow emission. The velocity ranges chosen for the \super{12}CO(2-1) emission highlight at least six (labeled with letters in Figure \ref{fig:12cooutflows}) and up to eleven previously unknown outflows in the central region of the MonR2 cluster. The orientations of the outflows are indicated by black arrows in Figure \ref{fig:12cooutflows}. IRS3 and IRS5 are both associated with bright and highly collimated bipolar outflows (left panel of Figure \ref{fig:12cooutflows}.) The CO emission around IRS5 is aligned along the northeast to southwest axis, exhibiting a large degree of collimation (outflow B). The redshifted component of the outflow appears between 26 and 40 km~s\super{-1} and features two separate $12\sigma$ and $15\sigma$ peaks. These knots, or "bullets", are commonly observed in low-mass outflows such as e.g. L1448-mm and HH211 \citep[see][]{guilloteau92,hirano06,palau06,lee07,hirano10} and may represent outbursts due to disk variability caused by episodic, unsteady accretion \citep{qiu09,lee10}. Another redshifted emission peak is detected further to the northeast (+13'', +40'') and appears more prominently in primary-beam corrected maps, where it is well-aligned with the rest of the redshifted lobe. On these grounds we calculate the physical parameters of the outflow in Section \ref{subsubsec:12coproperties} assuming all three bullets belong to the redshifted lobe. Its blueshifted counterpart appears for both velocity ranges between -16 and -5 km~s\super{-1} and -5 and 1.9 km~s\super{-1}, and it is less extended spatially. IRS5 is located at the geometric center of this outflow and is therefore likely its driving source (see Section$\,$\ref{subsubsec:12cocompwithIR}). Similarly, IRS3 is associated with the geometric center of  a bipolar outflow along the northeast to southwest axis (outflow A). The red- and blueshifted components of this outflow are clearly seen in both panels of Figure~\ref{fig:12cooutflows}. The blue lobe features two separate knots and is spatially more extended.

The right panel of Figure \ref{fig:12cooutflows} presents additional outflows identified in the 21 to 26 km~s\super{-1} and -5 to 1.9  km~s\super{-1} velocity ranges for redshifted and blueshifted emission, respectively. A prominent large scale redshifted lobe is present along the northwest-southeast direction in the vicinity of IRS5 (outflow D). This outflow lacks an obvious blueshifted counterpart. We propose that the most likely scenario is that the corresponding blue lobe could be located outside the field of view of our observations. The extended redshifted lobe features three approximately equally-spaced peaks. The angular separation between the knots is roughly 20'', giving a projected separation of approximately  $1.5\times10^4$~AU~$\simeq 0.08$~pc. If these bullets are associated with periodic outbursts, this separation between knots implies time-scales of $\sim$5000 yrs between bursts, which are consistent with those predicted by episodic accretion models in low-mass protostars \citep[see][]{baraffe10,zhu10}.
We also detect one red and two blue lobes with aligned elongation axes on either side of IRS1 (outflow E). Between IRS3 and IRS5, the image reveals the presence of two bright blueshifted condensations (C and F). The emission is strong for a wide range of velocities, lacks a well-defined directional axis and well-defined redshifted counterparts. Outflow C is also seen in the left panel of Figure$\,$\ref{fig:12cooutflows} for the velocity range from -16 to -5$\,$km$\,$s$^{-1}$. Finally, we detect a number of relatively bright structures located beyond the edge of the primary beam, which we tentatively assign to outflows according to their association with sources detected in the near-IR across the cluster (Section$\,$\ref{subsubsec:12cocompwithIR}). These outflows can be found toward the northern edge of the image, west of IRS5, and due north of IRS3 (right panel of Figure$\,$\ref{fig:12cooutflows}). They are unlikely to be density enhancements of the large-scale outflow observed by \citet{wolf90} because there are blue- and redshifted lobes present in both the eastern and western parts of the region, while the large CO outflow is blueshifted towards the northwest and redshifted in the southeast \citep{torrelles83}.

\subsubsection{Comparison with IR sources}
\label{subsubsec:12cocompwithIR}

\begin{figure*}
\begin{center}
\includegraphics[scale=0.6,trim = 0mm 20mm 10mm 0mm, clip]{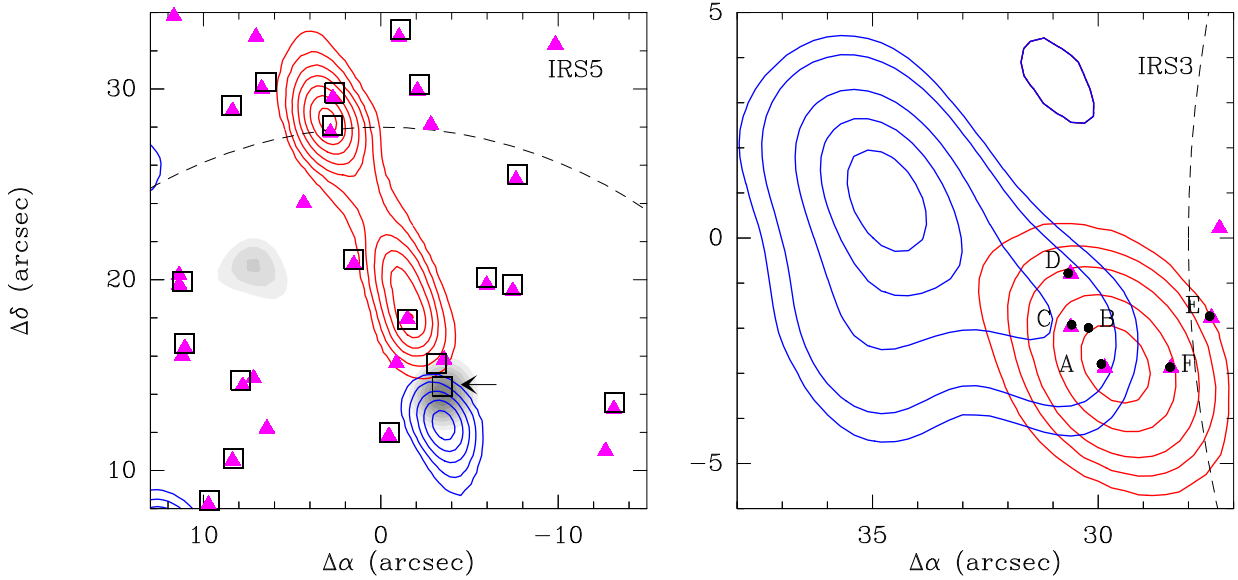} 
\caption{Moment-zero maps of the CO(2-1) transition in MonR2 (colored contours) overlaid on the 1.3~mm continuum (greyscale). Left panel: A zoom-in on outflow B near IRS5 where the redshifted emission has been integrated over the 26-40~km~s$^{-1}$ velocity range ($\sigma \sim 0.15$~Jy/beam km~s$^{-1}$), and the blueshifted emission over the -16 to -5~km~s$^{-1}$ range ($\sigma \sim 0.14$~Jy/beam km~s$^{-1}$).  The first contour and step levels are $4\sigma$. The location of the candidate driving source is marked by a black arrow. Right panel: A zoom-in on outflow A near IRS3 where the redshifted gas appears for velocities comprised between 21 and 26~km~s$^{-1}$ ($\sigma \sim 0.60$~Jy/beam km~s$^{-1}$). The 3$\sigma$, 5$\sigma$ and 7$\sigma$ contours are shown; contour spacing is $3\sigma$ thereafter. The blueshifted counterparts appear between -5~km~s$^{-1}$ and 1.9~km~s$^{-1}$ ($\sigma \sim 0.41$~Jy/beam km~s$^{-1}$). Here the 3$\sigma$, 5$\sigma$ and 10$\sigma$ contours are shown; contour spacing is $5\sigma$ thereafter. Magenta triangles indicate the positions of the IR sources observed with NICMOS by \citet{andersen06}, black squares the stellar population described in \citet{carpenter97} , and black dots and letters the IR sources identified by \citet{preibisch02}. The dashed circle indicates the primary beam of the observations, $\sim$57'' at 1.3~mm. The maps were made using uniform weighting.
\label{fig:12cooutflowszoom} }
\end{center}
\end{figure*}

The inner region of the MonR2 stellar cluster, approximately the field of view of our observations, is known to harbor a centrally peaked gas density distribution where the H$_2$ volume density, $n$(H$_2$), peaks toward the position of the IRS1 and IRS2 sources \citep[see Figures$\,$8 and 9 in][]{choi00}. In addition, surveys in the infrared \citep{carpenter97, andersen06} and in the X-rays \citep{kohno02,nakajima03,preibisch02} have also revealed a dense population of embedded low-mass stars. Here we compare the cluster of molecular outflows described in Section \ref{subsubsec:12comorphology} with the IR sources found by \citet{carpenter97}, \citet{preibisch02} and \citet{andersen06}, in order to identify the possible driving sources that power the CO outflows detected in MonR2.  

In Figure \ref{fig:12cooutflowszoom}, we present zoom-in maps of the CO outflows detected toward IRS5 and IRS3 (outflows B and A in Figure \ref{fig:12cooutflows}.) From the left panel of Figure \ref{fig:12cooutflowszoom}, we find that the collimated CO outflow toward IRS5 is likely associated with the star detected in the near-IR only by \citet{carpenter97} and which falls close to the location of the IRS5A source identified in the VEX data, the most massive star in the IRS5 sub cluster (its envelope mass is the largest one measured in the cluster; see Figure \ref{fig:VEXCom} and Table~$\,$\ref{tab:COMVEXcontinuum}). The candidate source is marked by an arrow in Figure \ref{fig:12cooutflowszoom}.

The right panel of Figure \ref{fig:12cooutflowszoom} shows the presence of a stellar subcluster in IRS3 (sources A-F) as detected from IR data \citep[][]{preibisch02}. As revealed by K-band images \citep{preibisch02}, the massive star IRS3 B presents an elongated nebulous feature that points toward the north-east with a P.A. of $\sim$50$^\circ$. This feature could be the near-IR counterpart of the blue-shifted lobe of the IRS3 CO outflow. The stellar mass of the IRS3 B source is $\sim$8-12$\,$M$_\odot$, as reported by \citet{preibisch02}. On the other hand, IRS3 A has a more extended IR nebulosity in the south-west direction. Although our CO data do not have sufficient spatial resolution to pinpoint whether IRS3 A or B is the driving source, the orientation of these nebulous features provides clues. Blue-shifted lobes pointing in the observer's direction suffer less from extinction than the red-shifted lobes and the former are typically detected in the IR. Since the direction of the IR nebulosity associated with IRS3 B \citep[i.e. the microjet reported by][]{preibisch02} matches the orientation of the blueshifted lobe observed with the SMA, IRS3 B is the more likely driving source of this outflow.

Besides the IRS5 and IRS3 CO outflows, the large-scale redshifted outflow labeled D in the right panel of Figure \ref{fig:12cooutflows} is closely aligned with an IR star detected in the NW corner of the image. This star is likely to be the outflow driver, consistent with the hypothesis that the blueshifted counterpart lies outside of the field of view to the northwest. The elongation of the knots on either side of IRS1 point to the location of the massive star IRS1 SW. Since this object is the exciting source of the MonR2 HII region (and it is thus at a late stage of evolution), it is possible that a low-mass companion is responsible for the CO molecular outflow labeled E. For the blue-shifted condensation close to IRS3 visible between -5 and -1.9~km~s$^{-1}$ (outflow F), IR observations reveal a nearby star aligned with the elongation axis, which could be its driving source. 
The exciting object of the blueshifted outflow C is less clear. The elongation axis suggests that one of the stars in the surroundings of IRS2 could be the driving source of outflow C, as indicated by the arrow in Figure \ref{fig:12cooutflows}. For the other five additional outflow candidates (most of them located outside of the primary beam of the observations), their emission peaks appear associated with IR sources. The presence of IR stars at approximately the outflow geometric centers suggests that these outflows are true detections.

\subsubsection{Physical properties}
\label{subsubsec:12coproperties}

In this section, we estimate the physical properties of the six brightest CO outflows detected in MonR2 which present robust identifications (i.e. outflows A to F; see Figure~\ref{fig:12cooutflows}).
\super{13}CO(2-1) emission was detected only in the blueshifted lobes of outflows A and B. The  \super{12}CO emission from other outflow components without  \super{13}CO detection was assumed to be optically thin.
For the blue lobes of outflows A and B, where \super{13}CO(2-1) emission is detected, the optical depths are derived using Equation~1 in \citet{roberts10} assuming a $^{13}$C/$^{12}$C isotopologue ratio of $\sim$1/69 \citep[see][]{wilson99}. The derived values are 3.7 for the blue lobe of outflow A, and 11 for the blue lobe of outflow B. From this, the masses of these two lobes are higher by a factor of $\sim$4 and 11, respectively. The correction factor is calculated as $\tau/(1-\exp \sqb{-\tau})$, where $\tau$ is the optical depth.

We apply the method described in \citet{garden91} and \citet{scoville86} to the CO(2-1) transition:
\begin{equation}
N_{\text{CO}} = 1.08 \times 10^{13} (T_{ex} + 0.92) \exp\bra{\frac{16.6}{T_{ex}}} \int T_B dv \,\,\, cm^{-2}
\end{equation}
\label{eq:columndensity}
Here 16.6~K is the upper level energy of the transition, $dv$ is the velocity interval in km~s\super{-1}, and $T_{ex}$ and $T_B$ are the excitation temperature and brightness temperature, respectively. The outflow mass, momentum, energy, dynamical age, outflow rate and momentum rate are then given by:
\begin{eqnarray}
M &=& d^2 \sqb{\frac{\text{H}_2}{\text{CO}}} \mu_{\rm{g}} m(\rm{H}_2) \int_\Omega N_{\text{CO}}(\Omega') d\Omega' \\
P &=& Mv \\
E &=& \frac{1}{2}Mv^2 \\
t_{\rm{dyn}} &=& \frac{\text{Length}}{v_{\rm{max}}} \\
\dot{M}_{\rm{out}} &=& \frac{M}{t_{\rm{dyn}}} \\
F_{\rm{out}} &=& \frac{P}{t_{\rm{dyn}}} \\
\end{eqnarray}
\label{eq:physparams}
We assumed a CO to H\sub{2} abundance of [CO/H$_2$]=10\super{-4} and a distance $d$ of 830~pc \citep{herbst76}. $\mu_{\rm{g}}$ is the mean atomic weight of the gas ($\mu_{\rm{g}}$=1.41) and $m$(H$_2$) is the mass of the hydrogen molecule.
The momentum and energy associated with each velocity channel $v$ in the ranges presented in Figure \ref{fig:12cooutflows} are added up to give total estimates for each outflow. The physical parameters of the CO outflows are reported in Table \ref{tab:physparams}. Since the shocked CO gas in outflows typically presents temperatures ranging from 30~K to 60~K \citep[e.g.][]{beuther02,arce10}, we provide estimates of the physical parameters based on this temperature range.
The masses we obtain are in the 10\super{-4} to 10\super{-2} M\sub{$\Sun$} range.
The momenta are of order 10\super{-3} - 10\super{-1} M\sub{$\Sun$}~km~s\super{-1}, while the energies are of order $\sim$0.01-3.0 M\sub{$\Sun$}~km~\super{2}~s\super{-2}. From these parameters, the dynamical ages of the outflows are found to lie between $\sim$1000 and $\sim$6000~yr, with corresponding outflow mass loss rates lying between $\sim$0.1-17$\times$10\super{-6} M\sub{$\Sun$} yr\super{-1}.

The outflow mass, momentum and energy are smaller than those from protostars of similar luminosities \citep{zhang01,zhang05,beuther02}. However, besides the optical depth correction, several additional effects could contribute to increasing our estimates:
i) outflow gas at velocities near the ambient clump velocity \citep{arce10}; ii) projection effects \citep{arce10}; iii) 50\% of the gas in an outflow could be in atomic form \citep{reipurth01}; iv) presence of high-velocity gas below the sensitivity of our observations \citep{dunham14}; and v) the values presented here represent lower limits since the spatially extended emission in the outflow, which can be a large fraction of the total, may not be recovered in the SMA observations (e.g. Zhang et al., 2000; Qiu et al., 2009).

\begin{deluxetable}{ccccccc}
\tablewidth{0pt}
\tabletypesize{\small}
\tablecaption{Physical parameters of the CO outflows detected toward MonR2 with the SMA.}
\tablehead{
\colhead{Lobe} & \colhead{Mass} & \colhead{Momentum} & \colhead{Energy} & \colhead{$t_{\rm dyn}$} & \colhead{$\dot{M}_{\rm out}$} & \colhead{$F_{\rm out}$} \\
&(10\super{-3} M\sub{$\Sun$}) & (10\super{-2}$~$M\sub{$\Sun$} km~s\super{-1}) & (M\sub{$\Sun$} km\super{2} s\super{-2}) &  (10$^3$ yr) & (10\super{-6} M\sub{$\Sun$} yr\super{-1}) & (10\super{-5} M\sub{$\Sun$} km s\super{-1} yr\super{-1})
} 
\startdata
\sidehead{A - IRS3 outflow}
Blue & 20.7-30.3 & 25.2-38.0 & 1.7-2.5 & 1.8 & 11.5-17.0 & 14.1-20.7 \\
Red & 1.8-2.6 & 2.8-4.2 & 0.3-0.4 & 2.0 & 0.9-1.3 & 1.4-2.1\\
\hline
\sidehead{B - IRS5 outflow}
Blue & 4.4-5.5 & 6.6-8.8 & 0.55-0.77 & 1.2 & 3.6-4.6 & 5.5-7.3\\
Red & 1.6-2.4 & 3.4-5.1 & 0.4-0.6 & 4.9 & 0.3-0.5 & 0.7-1.0 \\
\hline
\sidehead{C - Westernmost isolated outflow between IRS3 and IRS5}
Blue & 2.6-3.9 & 3.1-4.6 & 0.20-0.29 & 1.5 & 1.7-2.6 & 2.1-3.1 \\
\hline
\sidehead{D - Large-scale redshifted lobe west of IRS5}
Red & 7.2-10.6 & 8.3-12.2 & 0.5-0.7 & 6.2 & 1.2-1.7 & 1.3-2.0 \\
\hline
\sidehead{E - IRS1 outflow}
Blue & 1.6-2.7 & 1.5-2.4 & 0.07-0.1 & 3.9 & 0.4-0.7 & 0.4-0.6 \\
Red & 0.2-0.3 & 0.24-0.30 & 0.01-0.02 & 2.0 & 0.11-0.13 & 0.12-0.15\\
\hline
\sidehead{F - Easternmost isolated outflow between IRS3 and IRS5}
Blue & 1.3-1.8 & 1.2-1.8 & 0.06-0.09 & 2.1 & 0.6-0.9 & 0.6-0.8 \\
\enddata
\tablecomments{Estimates in columns 2, 3, 4, 5 and 6 are provided for a gas temperature range of 30 to 60~K. For the red lobe of B, the IRS5 outflow, the estimates include the knot up far to the northeast of IRS5 toward position (13$"$, 40$"$) in the left panel of Figure~\ref{fig:12cooutflowszoom} (best visible in the primary-beam corrected map). Without this additional bullet, the mass, momentum and energy estimates are smaller by about 10\%.}
\label{tab:physparams}
\end{deluxetable}

\subsubsection{Spectral Energy Distribution (SED) fitting of IRS5}
\label{subsubsec:sedfitting}

\begin{figure*}
\begin{center}
\includegraphics[scale=0.8]{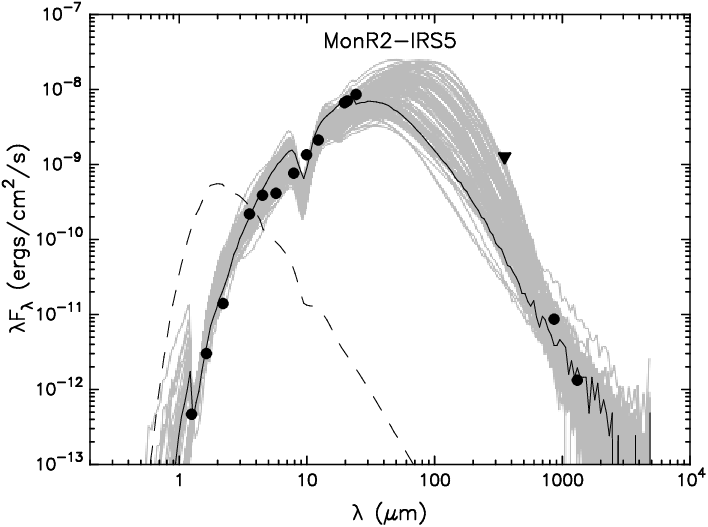} 
\caption{SED fitting of the IRS5 source. Filled circles show the fluxes measured with the SMA (Table~$\,$\ref{tab:COMcontinuum}) and those observed in the near- and mid-IR, which are available in the literature \citep[see][]{hackwell82,carpenter97,kraemer01,mueller02,gutermuth09,dewit09}. Filled triangle indicates an upper limit to the measured flux. The black line shows the best fit to the data while the grey lines indicate the 100 best fits to the IRS5 SED. The dashed line shows the emission arising from the star's photosphere. The results of the SED fitting provide a stellar mass and luminosity of $\sim$7$\,$M$_\odot$ and $\sim$300$\,$L$_\odot$ respectively for the IRS5 source. \label{fig:SEDIRS5}}
\end{center}
\end{figure*}

To obtain further information about the physical properties of IRS5 and about its evolutionary stage, we have built the SED of this source by using the fluxes measured by the SMA in compact configuration (Table~$\,$\ref{tab:COMcontinuum}) and those observed in the near- and mid-IR, available in the literature \citep{hackwell82,carpenter97,kraemer01,mueller02,gutermuth09,dewit09}. These values have then been fitted by means of the radiative transfer models of \citet{robitaille06}. Note that like most of the IR observations, our data from the SMA compact configuration do not resolve the IRS5 A and B sub-sources identified in VEX configuration (see Figure~\ref{fig:VEXCom}) and therefore the SED fitting has been performed for source IRAS5 A+B. These results are reported in Figure~\ref{fig:SEDIRS5}. The best fit to the IRS5 SED provides a stellar mass of $\sim$7$\,$M$_\odot$ and a stellar luminosity of $\sim$300$\,$L$_\odot$. This stellar luminosity is consistent with that previously derived by \citet{hackwell82} from 10-20 micron colors. The source presents a stellar age of $\sim10^4$$\,$yr, which is consistent with the idea of a young stellar object that is still accreting gas. We however point out that this estimate is only approximate since the stellar evolutionary tracks used in the models of \citet{robitaille06} are from canonical non-accreting pre-main-sequence stars \citep{bernasconi96, siess00}. \citet{robitaille06} indeed prefer to use an evolutionary classification into Stage 0/I, II and III sources instead of using stellar ages directly. Our source would therefore fall in Stage 0/I after comparison to the SED shapes shown in Figure 7 of \citet{robitaille06}.
The IRS5 source is deeply embedded in the cloud with derived circumstellar extinction of $A_v\sim$50 mag, roughly consistent with the extinction inferred from the H$_2$ column densities measured toward IRS5 (i.e. $\sim$20-40 magnitudes; see Section~\ref{subsubsec:molecular gas properties}). The estimated disk mass is high ($\sim$0.2$\,$M$_\odot$), supporting the idea that IRS5 is relatively young.
The derived disk accretion rate is $\sim$10$^{-6}$~M\sub{$\Sun$}~yr\super{-1},  which is similar to the outflow rate calculated for outflow B (see Table~\ref{tab:physparams}). We note that the disk parameters derived from the SED fitting should be taken with caution, due to the multiple nature of the IRS5 source and the relatively low-angular resolution (arcsecond scales) of the IR observations.
The rich molecular line emission detected toward IRS5 (see Section~\ref{subsec:molecular}) also indicates that IRS5 is at an early stage of evolution characterized by a hot-core-like chemistry.   

\subsection{Molecular line emission from Other Molecules}
\label{subsec:molecular} 

\subsubsection{Morphology of the Molecular Gas toward MonR2}
\label{subsubsec:molecular gas morphology}

\begin{figure*}
\begin{center}
\includegraphics[scale=0.67,angle=270]{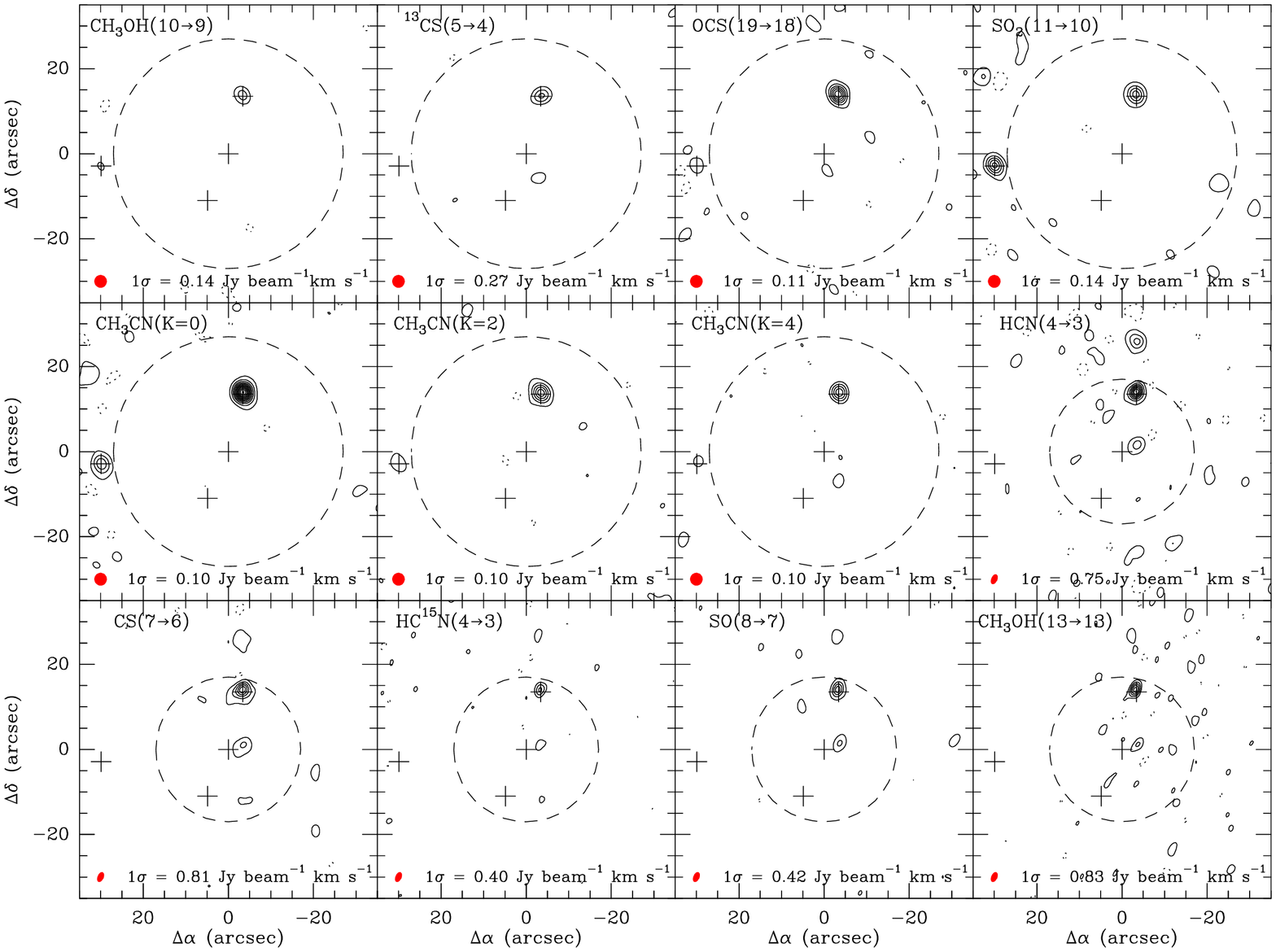} 
\caption{Images of the MonR2 region for different spectral lines, indicated in the upper left-hand corner of each panel. The maps show the integrated emission from 5 to 15~km~s\super{-1}, with each line's central velocity set at 10~km~s\super{-1}. As before, crosses mark the positions of the main IRS sources in the region and dotted contours indicate the negative $3\sigma$ level. The first contour and step level are $3\sigma$, except for OCS($19 \rightarrow 18$), SO\sub{2}($11 \rightarrow 10$), CH\sub{3}CN ($K=0$ and $K=2$), HCN($4 \rightarrow 3$), CS($7 \rightarrow 6$) and SO($8 \rightarrow 7$), where the contour spacing is $6\sigma$. The beam size is shown as a red ellipse in the lower left corner: $2.99''\times2.98'',$ P.A.= $86.6^\circ$ at 1.3~mm, and $2.44''\times1.57'',$ P.A.= $114.7^\circ$ at 0.85~mm. The dashed circle indicates the primary beam of the observations, $\sim57''$ at 1.3~mm, and $\sim35''$ at 0.85~mm. The maps were made using natural weighting.
\label{fig:12Linefig} }
\end{center}
\end{figure*}

In addition to the continuum emission 2~mm and 0.85~mm and to the $^{12}$CO $J$=2$\rightarrow$1 and $^{13}$CO $J$=2$\rightarrow$1 rotational lines, our SMA data report the detection of several molecular line transitions towards the MonR2 star forming region. The detected rotational lines are shown in Table \ref{tab:mollines}, and include lines from molecular species typical of hot core chemistry such as CH\sub{3}CN, CH\sub{3}OH and sulphur-bearing molecules such as SO\sub{2}, CS and OCS. The integrated intensity images of some representative lines are presented in Figure \ref{fig:12Linefig}. 

From Figure \ref{fig:12Linefig}, we find that the molecular emission in the MonR2 region is concentrated toward the IRS3 and IRS5 sources. IRS3, with a luminosity of 1.5$\times$10\super{4} L\sub{$\Sun$} \citep{henning92,giannakopoulou97}, is expected to show a rich chemistry as a consequence of the evaporation of the mantles of dust grains as revealed by the detection of warm H\sub{2}O, SO$_2$ and CH$_3$OH molecular gas \citep[excitation temperatures of 110-125~K for SO$_2$ and CH$_3$OH and of 250~K for H\sub{2}O; see][]{vandertak03,boonmanvandishoeck03,boonman03}. However, only lines from SO\sub{2} and the $K$=0 to $K$=5 transitions of CH\sub{3}CN are measured toward this source. 
This is probably due to the fact that IRS3 falls outside of the primary beam of our observations, preventing the detection of fainter molecular lines. New high angular resolution observations centered on IRS3 are needed to study in detail the chemistry of the molecular gas toward this source.

Figure \ref{fig:12Linefig} also shows that most of the molecular emission in our SMA images arises from IRS5 and thus, besides IRS3, this source appears as the most chemically active object in the region. This confirms IRS5's nature as a very young object with a chemically rich envelope. The molecular line emission toward IRS5 is very compact and its line profiles show a single Gaussian component with line widths ranging from 5 to 7 km~s\super{-1}. We do not detect any broad velocity emission above the $3 \sigma$ level in our images that could be associated with the high-velocity CO outflow reported in Section \ref{subsec:12co}. This suggests that the molecular line emission detected toward IRS5 likely arises from the envelope surrounding this source, instead of from outflowing shocked gas.

We note that the other two molecular condensations seen toward the north and south of IRS5 in HCN, CS and SO, are side lobes generated by the limited UV plane coverage of our 0.85~mm observations. The presence of significant sidelobes in these images is due to a dynamic range problem as a result of the bright emission of these lines.

\subsubsection{Excitation Temperatures, Column Densities and Molecular Abundances toward MonR2 IRS5 and IRS3}
\label{subsubsec:molecular gas properties}

\noindent \emph{Methyl cyanide}

{\rotate{
\begin{deluxetable}{lcccccc}
\tabletypesize{\footnotesize}
\tabcolsep 3pt 
\tablewidth{0pt}
\tablecaption{Derived parameters of the molecular line emission measured toward IRS5.}
\tablehead{
\colhead{Line} & \colhead{$\nu$ (GHz)} & \colhead{$E_{u} (K)$}  & \colhead{Line Flux (Jy/beam km~s$^{-1}$)} & \colhead{$v_{LSR}$ (km~s$^{-1}$)} & \colhead{$\Delta v$ (km~s$^{-1}$)} & \colhead{$I_{peak}$ (Jy/beam)}
}
\startdata
CH\sub{3}CN ($12 \rightarrow 11, K = 0$) & 220.74726 & 69 & $6.21\pm0.05$ & $9.34\pm0.08$ & $7.3\pm0.2$ & $0.801\pm0.018$ \\
CH\sub{3}CN ($12 \rightarrow 11, K = 1$) & 220.74301 & 76 & $3.22\pm0.05$ & $9.4\pm0.3$ & $6.1\pm0.4$ & $0.492\pm0.021$ \\
CH\sub{3}CN ($12 \rightarrow 11, K = 2$) & 220.73026 & 98 & $4.28\pm0.06$ & $8.89\pm0.18$ & $6.9\pm0.4$ & $0.583\pm0.020$ \\
CH\sub{3}CN ($12 \rightarrow 11, K = 3$) & 220.70902 & 133 & $4.85\pm0.06$ & $8.75\pm0.14$ & $7.4\pm0.3$ & $0.619\pm0.021$ \\
CH\sub{3}CN ($12 \rightarrow 11, K = 4$) & 220.67929 & 183 & $1.87\pm0.05$ & $8.89\pm0.21$ & $6.5\pm0.6$ & $0.268\pm0.020$ \\
CH\sub{3}CN ($12 \rightarrow 11, K = 5$) & 220.64108 & 248 & $1.18\pm0.04$ & $9.20\pm0.15$ & $4.5\pm0.4$ & $0.247\pm0.020$ \\
CH\sub{3}CN ($12 \rightarrow 11, K = 6$) & 220.59442 & 326 & $1.31\pm0.06$ & $9.00\pm0.30$ & $6.5\pm0.7$ & $0.190\pm0.022$ \\
CH\sub{3}CN ($12 \rightarrow 11, K = 7$) & 220.53932 & 419 & $\leq 0.14$ & - & - & $\leq 0.05$ \\
SO\sub{2} (11$_{1,11}$$\rightarrow$10$_{0,10}$) & 221.96521 & 60 & $4.47\pm0.05$ & $9.09 \pm 0.14$ & $6.1 \pm 0.3$ & $0.646 \pm 0.020$\\
OCS ($19 \rightarrow 18$) & 231.06098 & 111 & $5.35 \pm 0.07$ & $8.95 \pm 0.09$ & $6.5 \pm 0.2$ & $0.78 \pm 0.03$ \\
\super{13}CS ($5 \rightarrow 4$) & 231.22099 & 33 & $3.14 \pm 0.15$ &  $9.22 \pm 0.23$ & $6.2 \pm 0.5$ & $0.48 \pm 0.06$\\
CH\sub{3}OH (10$_{2,9} \rightarrow$9$_{3,6}$ A$^{-}$) & 231.28111 & 165 & $4.07 \pm 0.07$ & $8.76 \pm 0.11$ & $5.8 \pm 0.2$ & $0.66 \pm 0.03$ \\
CH\sub{3}OH ($13_{1,12} \rightarrow 13_{0,13}$ A) & 342.72980 & 228 & $20.6 \pm 0.3$ & $9.07 \pm 0.12$ & $6.1 \pm 0.3$ & $3.16 \pm 0.16$ \\
CS ($7 \rightarrow 6$)  & 342.88300 & 66 & $42.8 \pm 0.23$ &$8.59 \pm 0.01$ & $3.72 \pm 0.03$ & $10.8 \pm 0.1$\\
HC\super{15}N ($4 \rightarrow 3$) & 344.20032 & 41 &$6.8 \pm 0.3$ & $8.21 \pm 0.24$ & $5.2 \pm 0.6$ & $1.24 \pm 0.16$\\
SO ($8 \rightarrow 7$) & 344.31061 & 88 &  &  &  &  \\
\quad Component 1 &  &  & $15.8 \pm 0.2$ & $7.3 \pm 0.7$ & $6.4 \pm 0.7$ & $2.3 \pm  0.1 $\\
\quad Component 2 &  &  & $3.42 \pm 0.18$ & $9.7 \pm 0.7$ & $4.4 \pm 0.7$ & $0.7 \pm  0.1 $\\
HCN ($4 \rightarrow 3$) & 354.50548 & 43 & $60.2 \pm 0.7$ & $8.18 \pm 0.03$ & $5.4 \pm 0.1$ & $10.4 \pm 0.4 $\\
\hline
\enddata
\tablecomments{ The errors on the line flux were calculated as $\sigma \sqrt{\delta v \times \Delta v}$, where $\delta v$ is the velocity resolution in the spectrum and $\Delta v$ is the linewidth (column 6 in this Table). Since the CH\sub{3}CN ($K $= 7) transition was not detected, the upper limit on its peak intensity is given by the 3$\sigma$ level in the spectrum, and the upper limit on the integrated intensity as 3$\sigma \sqrt{\delta v \times \Delta v}$. The SO ($8 \rightarrow 7$) transition presented with a double Gaussian profile corresponding to two velocity components.} 
\label{tab:mollines}
\end{deluxetable}
}}

\begin{figure*}
\begin{center}
\includegraphics[scale=0.8]{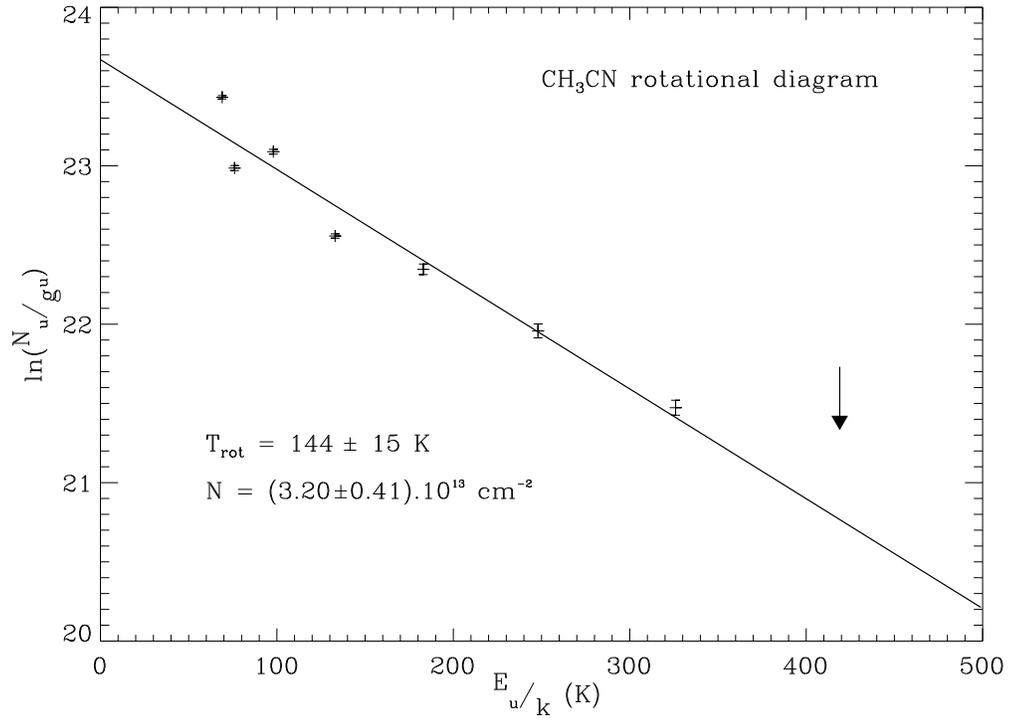} 
\caption{Methyl cyanide rotational diagram for the $K = 0$ to $K = 6$ ladder detected toward IRS5. The data are shown as black plus signs with corresponding $1\sigma$ error bars, and the linear fit as a black line. The upper limit derived for the $K = 7$ transition is shown as a black arrow.
\label{fig:rotationaldiag} }
\end{center}
\end{figure*}

Within our dataset toward MonR2, we have detected seven rotational transitions of the methyl cyanide molecule (CH\sub{3}CN.) This molecule is commonly used to derive the excitation temperature of warm/hot gas in star-forming regions. In Figure~\ref{fig:12Linefig}, we report the images for the CH\sub{3}CN $K = 0$, $K = 2$ and $K = 4$ transitions detected with the SMA. The CH$_3$CN line spectra at the position of the IRS5 continuum peak were extracted from the image cube after primary-beam correction. The line parameters listed in Table~\ref{tab:mollines} were derived from the fitting of the CH\sub{3}CN $K = 0$ to $K = 6$ transitions with single component Gaussian line profiles. The $K = 0$ and $K = 1$ transitions are very close in frequency and hence the accuracy in the fits suffers from partial merging of the lines. The excitation temperature of the CH\sub{3}CN gas was then estimated using the rotational diagram method \citep[][see also Appendix A]{zhang98,goldsmith99}. 

The rotational diagram of the CH\sub{3}CN molecule, derived by following the method described in the Appendix, is shown in Figure \ref{fig:rotationaldiag}. From the slope of the linear least squares fit to the data, we estimate an excitation temperature for the CH\sub{3}CN gas of $144\pm15$~K. This value is consistent with typical gas temperatures found in massive hot core sources such as Cepheus A HW2 \citep[see][]{martin-pintado05,jimenez-serra09}. The derived total column density of CH\sub{3}CN toward IRS5 is $(3.2\pm0.4)\times10^{13}$~cm\super{-2}. 

Although IRS3 falls outside the primary beam of the SMA observations, we can provide an estimate for the excitation temperature of the gas toward this source from the measured CH$_3$CN line emission. The derived excitation temperature is $126\pm22$~K.This temperature is consistent with those measured by \citet{vandertak03} from single-dish observations of several SO$_2$ and CH$_3$OH lines ($\sim$110~K to 125~K). We note that we provide no additional details for the CH$_3$CN rotational diagram calculated for IRS3 because the K $>2$ lines are detected only at the $\sim 3 \sigma$ level (the source falls at the edge of the primary beam).

\noindent \emph{Other molecular species}

The column densities of the other molecular species detected toward IRS5 are estimated by assuming LTE and optically thin emission. In the case of HCN, the emission is not optically thin because HC$^{15}$N is detected (see Table \ref{tab:mollines}). Therefore, we correct for optical depth (see the expression for the correction factor in Section$\,$\ref{subsubsec:12coproperties}) based on an isotopic ratio of $1/388$ \citep{wilson99}. The derived optical depth for HCN is $\tau$$\sim$47. We use Equation \ref{eq:Nugu} to determine the values of $N_u/g_u$ for every rotational transition, and then calculate the total column density, $N_{tot}$ of the molecular species by reformulating \ref{eq:rotdiageq} into:
\begin{equation}
N_\text{tot} = \frac{N_u}{g_u} Q(T_\text{rot}) e^{E_u/k T_\text{rot}}.
\label{eq:coldensity}
\end{equation}
The values of $S\mu^2$ and $Q(T_\text{rot} \simeq 150$~K) are taken from the Cologne Database for Molecular Spectroscopy \citep[CDMS;][]{mueller05}. We estimate the molecular abundances, $\chi$, by matching the beams of the 1.3 mm and 0.85 mm continuum maps with those of the line maps at 1.3~mm and 0.85~mm, respectively. The H$_2$ column density toward IRS5 is calculated from the measured 1.3 mm and 0.85 mm peak fluxes of IRS5 (Table~$\,$\ref{tab:COMcontinuum}) following the method in \citet{enoch06}. We assume a gas-to-dust ratio of 100. The derived H$_2$ column densities toward this source are $1.7\times10^{22}$~cm\super{-2} from 1.3~mm and $3.8\times10^{22}$~cm\super{-2} from 0.85~mm. These values are derived from the peak fluxes at different angular resolutions to match the peak flux of the molecular lines. The resulting molecular column densities and abundances are reported in Table~\ref{tab:coldens}. 

{\rotate{

\begin{deluxetable}{lccccccccc}
\tabletypesize{\footnotesize}
\tablewidth{0pt}
\tablecaption{Molecular column densities and abundances measured toward IRS5 with the SMA. }
\tablehead{
\colhead{} & \colhead{CH\sub{3}CN} & \colhead{SO\sub{2}} & \colhead{OCS} & \colhead{\super{13}CS} & \colhead{CH\sub{3}OH} & \colhead{CS} & \colhead{HC\super{15}N} & \colhead{SO} & \colhead{HCN} 
}
\startdata
$N$ ($\times 10^{15}$cm\super{-2}) & $0.032\pm0.004$ & $1.23\pm0.11$ & $1.31\pm0.08$ & $0.031\pm0.003$ & $24.5\pm2.0$ & $46.8\pm0.004$ & $0.027\pm0.001$ & $1.12\pm0.02$ & $9.87\pm0.001$ \\
$\chi$ ($\times 10^{-8}$) & 0.2 & 7.3 & 7.7 & 0.2 & 150 & 1.2 & 0.07 & 3.0 & 26.3 \\
\enddata
\label{tab:coldens}
\end{deluxetable}

}}

\section{Discussion}
\label{sec:discussion}

\subsection{Outflow feedback in the MonR2 star forming cluster}
\label{subsec:feedback}

The presence of multiple outflows from a cluster of low- to intermediate-mass stars may have important implications in cluster and star formation, since outflows can inject large amounts of turbulence into the surrounding environment \citep[][]{li06,nakamura07,carroll09}. In this section we discuss the impact of the newly discovered  cluster of compact CO outflows in the center of the MonR2 clump, comparing with the one due to the already known large-scale outflow (\citealt{wolf90}).

The outflow feedback in a given environment depends on the outflow momentum as well as the mass and radius of the region being affected. The outflow momenta are presented in Table~\ref{tab:physparams}. \citet{ridge03} estimated a mass of $\sim$1800 M$_{\odot}$ for the MonR2 clump within a radius of $\sim$0.7 pc. \citet{walker90} found that the inner $\sim$0.22~pc of the clump contains $\sim$180 M$_{\odot}$. These values are consistent with the density profile of $r^{-1}$ found in the region by \citet{choi00}. The cluster of outflows presented in this paper is distributed within the inner $\sim$0.15 pc radius of the clump. Scaling the mass using the $r^{-1}$  density profile resulting from the estimates above, we obtain a mass of $\sim$75 M$_{\odot}$ for this region. In the following analysis, we will use $R_{\rm reg}$~=~0.15~pc and $M_{\rm reg}$~=~75~M$_{\odot}$ to evaluate the impact of the cluster of compact SMA outflows. In the case of the CO giant outflow, whose size is much larger ($\sim0.6-$7 pc; \citealt{wolf90}) $-$ and hence acts over a larger and more massive region$-$ we will use the values of the clump studied by \citet{ridge03}: $R_{\rm reg}$ = 0.7~pc and $M_{\rm reg}$ = 1800~M$_{\odot}$. In Table \ref{table-feedback} we summarize the physical properties of the two regions of interest, along with the outflow feedback parameters to be explained in the following subsections.

\subsubsection{Turbulent support of the MonR2 clump}
\label{subsec:turbulence}

In the absence of replenishment of the initial turbulence, the MonR2 clump is expected to collapse in a free-fall time, $t_{\rm ff}$\footnote{$t_{\rm ff}=\sqrt{\frac{3\pi}{32G\rho}}$, where $G$ is the gravitational constant and $\rho$ is the gas density, calculated as $\rho=\frac{3 M_{\rm reg}}{4\pi R_{reg}^3}$}. The free-fall time for the region with $R_{\rm reg}$ = 0.15~pc is estimated to be $t_{\rm ff}\sim$1$\times$10$^{5}$~yr, and for $R_{\rm reg}$= 0.7~pc it is $t_{\rm ff}\sim$2$\times$10$^{5}$~yr.
However, the estimated average age of the stellar cluster is larger than these values, $\sim$1 Myr \citep{carpenter97,andersen06}, which implies that a mechanism may have prevented the collapse of the MonR2 clump in its free-fall time, allowing the formation of stars. \citet{wolf90} already claimed that the large-scale CO outflow detected in the region could provide enough support to the whole clump. We evaluate here whether the population of CO outflows revealed by the SMA can also play a significant role. 

By following the analysis of \citet{rivilla13b}, we compare the rate at which the initial turbulence decays, $L_{\rm turb}$, with the rate at which the clump gains energy due to outflows, $L_{\rm gain}$. The former is given by $L_{\rm turb}=E_{\rm turb}/t_{\rm ff}$, with $E_{\rm turb}=(3/2)M_{\rm reg}\,\sigma_{\rm 1D}^{2}$, where $\sigma_{\rm 1D}$ is the observed one-dimensional velocity dispersion of the gas in the clump and $M_{\rm reg}$ the clump mass.
We evaluate the 1D velocity dispersion using the CS linewidth observed by \citet{tafalla97}. These authors showed that the linewidth decreases with the radius. From their Fig. 14 we consider an average value of linewidth of $\Delta v\sim $2.3 km s$^{-1}$ for radius $<$ 0.15~pc and $\sim$1.9 km s$^{-1}$ for radius $<$ 0.7~pc. For a 1D Maxwellian distribution of velocities, the relation between the linewidth and the one-dimensional velocity dispersion is $\Delta v=2\sqrt{2\ln2}\,\sigma_{\rm 1D}$, and hence we obtain $\sigma_{\rm 1D}\sim$0.96 km s$^{-1}$ and 0.81 km s$^{-1}$, respectively (Table \ref{table-feedback}).

On the other hand, $L_{\rm gain}$ is given by $\frac{1}{2} \dot M_{\rm core}\,\sigma_{\rm 3D}^{2}=\frac{\sqrt{3}}{2}F_{\rm out}\,\sigma_{\rm 1D}$, where $\dot M_{\rm core}$ is the mass set in motion by outflows per unit time and $F_{\rm out}$=$P_{\rm out}$/$t_{\rm dyn}$ is the outflow momentum rate.  
From the measured values of the CO molecular outflows (see Table~\ref{tab:physparams}) we obtain that $L_{\rm gain}/L_{\rm turb}\sim$0.3 (Table~\ref{table-feedback}). As shown in Figure~\ref{fig-outflow-feedback}, this value is similar to the ones found in several Perseus clumps \citep[][]{arce10}. 

We note that the values of the outflow parameters derived from the SMA data have likely been underestimated because of several reasons. \citet{arce10} suggested correction factors for the outflow momentum of 1.4 and 2 to account for inclination effects and for outflow gas at velocities near the ambient clump velocity. In Figure~\ref{fig-outflow-feedback} we have included a "correction vector" to account for these two effects, which yield a $L_{\rm gain}/L_{\rm turb}$ ratio of $\sim$0.9. This value is a lower limit, because the gas in an outflow could be in atomic rather than molecular form due to dissociative shocks (\citealt{reipurth01}), and because some high-velocity emission from the outflow may be below the sensitivity of our observations. 

Recently, \citealt{dunham14} quantified these corrections in a sample of single-dish outflow observations. Here we have used the mean corrections factors found in this study (factor of 41 and 93 for $p_{\rm outflow}$ and $F_{\rm outflow}$, respectively). Moreover, we have considered an additional factor of 2 to account for the presence of missing extended flux due to short-spacings not covered by the interferometer. Figure~\ref{fig-outflow-feedback} therefore also includes a second "correction vector" to account for all these effects. The uncorrected data from outflow observations by \citet{arce10} and \citet{stanke07} are shown for comparison. Considering all the corrections, the ratio $L_{\rm gain}/L_{\rm turb}$ for the SMA outflows would reach a value of $\sim$61 (Fig. \ref{fig-outflow-feedback}). This implies that even if only a fraction of the energy in the compact outfows is transformed into turbulent energy, the SMA cluster of outflows could help to maintain the turbulence in the MonR2 clump center. For the large-scale CO outflow, we obtain $L_{\rm gain}/L_{\rm turb}\sim$122 after applying all corrections except for the missing flux correction.

\begin{table}
\caption{Sizes and masses of the regions where the SMA cluster of compact outflows and the large-scale outflow are considered to act, along with the derived outflow feedback parameters.}             
\label{table-feedback} 
\vspace{0.5cm}  
\hspace{0cm}   
\centering          
\begin{tabular}{c| c c| c | c }     
\hline 
                  
Outflow &  $R_{\rm reg}$ & $M_{\rm reg}$ &  $L_{\rm turb}/L_{\rm gain}^{a}$ & $\sigma_{\rm 1D}$/$\sigma_{\rm gain}^{a}$\\
    &  (pc) & (M$_{\odot}$) &  &  \\
\hline  
SMA cluster    & 0.15 & 75    & 0.3 (0.9) [61]  &  0.01  (0.03) [0.9] \\
large-scale CO & 0.7  & 1800  & 0.7 (1.8) [122]  & 0.7   (2) [61] \\
\hline                 
\end{tabular}
\begin{list}{}{}
\item[$^{\mathrm{a}}$]{ The values in parentheses are corrected for gas at ambient velocities and projection effects following \citet{arce10}. The values in square brackets are corrected for gas at ambient velocities, projection effects, atomic/molecular gas fraction, sensitivity \citep[using the mean correction factors from][]{dunham14}. For the cluster of outflows observed with the SMA, the effect of missing flux was also corrected (see Section \ref{subsec:turbulence}).}
\end{list}
\end{table}

\subsubsection{Broadening of the linewidths induced by the population of the CO outflows}
\label{subsec:broadening}

As we already mentioned, \citet{tafalla97} found that the linewidths of the dense molecular tracer CS increase towards the center of the MonR2 region. They suggested that this is due to a combination of the broadening produced by the giant CO outflow and an additional turbulence source in the inner region. Since we have detected a cluster of young molecular CO outflows at the center of the MonR2 clump, we evaluate whether the outflow feedback can contribute to the broadening of the molecular linewidths. 

\begin{figure*}[h]
\begin{center}
\includegraphics[scale=0.5]{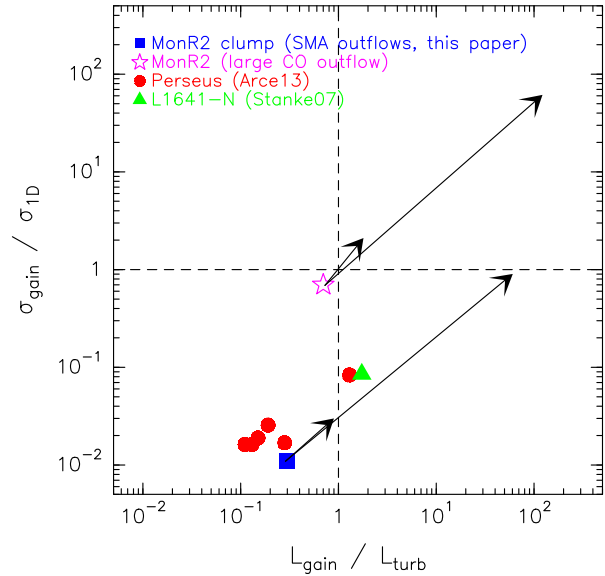} 
\caption{Outflow feedback from the MonR2 cluster of outflows. The blue filled square corresponds to the SMA outflows presented in this paper (correction factors not included). We compare this feedback with the impact of the giant outflow (magenta open star, \citealt{wolf90}) and several clusters of outflows from other star-forming regions. The short black arrows indicate the correction accounting for the underestimation of the outflow parameters following \citet{arce10}. Longer arrows account for the corrections found by \citet{dunham14} and an additional factor of 2 due to the lack of short-spacings.}
\label{fig-outflow-feedback}
\end{center}
\end{figure*}

Assuming that the total momentum of the SMA outflows is transferred to the gas, we obtain that the ratio between the velocity dispersion produced by outflows $\sigma_{\rm gain}=\frac{P_{\rm outflow}}{M_{\rm reg}}$ and the observed velocity dispersion $\sigma_{\rm 1D}$ is $\sigma_{\rm gain}$/$\sigma_{\rm 1D}$  $<<$ 1 (see Table \ref{table-feedback} and Figure \ref{fig-outflow-feedback}). However, considering the correction factors found by \citet{dunham14}, this value reaches $\sigma_{\rm gain}$/$\sigma_{\rm 1D}\sim$ 0.9. This implies that the turbulence injected by the cluster of CO outflows in the central region of MonR2 could be responsible for the observed velocity dispersion. In the case of the large-scale outflow, the corrected ratio  $\sigma_{\rm gain}$/$\sigma_{1D}$ is $>>$1 (see Table~\ref{table-feedback} and Figure~\ref{fig-outflow-feedback}), indicating that this agent may also contribute significantly to the velocity dispersion observed in the whole MonR2 clump.

The turbulence injected by the compact SMA outflows may also explain the difference in linewidth observed by \citet{tafalla97} between the center and the periphery of the clump. This difference is $\Delta v_{\rm diff}\sim$0.5 km s$^{-1}$ (\citealt{tafalla97}), which is equivalent to a velocity dispersion of $\sigma_{\rm diff}$ = 0.21 km s$^{-1}$. This value is lower than our corrected value for $\sigma_{\rm gain}\sim$ 0.85 km s$^{-1}$, suggesting that the broadening of the molecular linewidths in the center of the cluster may be produced by the compact CO outflows.

\subsection{IRS5: An intermediate- to high-mass star in the hot-core phase}
\label{subsec:irs5chem}

In Section$\,$\ref{subsec:molecular}, we have reported the detection of rich molecular line spectra toward the IR sources IRS3 and IRS5 in the MonR2 cluster. These spectra present several molecular line transitions from complex organic species, such as CH$_3$OH and CH$_3$CN, and from sulfur-bearing species, such as SO$_2$ or OCS. The high temperatures measured toward the innermost envelope regions around massive stars (of few 100 K) cause the thermal evaporation of the icy mantles of dust grains, chemically enriching their molecular environment \citep[these objects are commonly known as {\it hot cores}; see e.g.][]{charnley92,charnley97,viti04,wakelam04}. For IRS3, previous millimeter and infrared observations toward this source already noted a rich hot-core chemistry \citep[][]{boonmanvandishoeck03,vandertak03}. Although shocks are also expected to enhance the abundance of certain molecular species in star forming regions, the radiative transfer modelling study carried out by \citet{vandertak03} showed that their contribution is likely very little. IRS3 is a cluster of IR sources \citep[][]{preibisch02} and therefore the most massive objects in the region, IRS3 A and B (with derived star masses of $\sim$10-15$\,$M$_\odot$), are the best candidates for driving the hot-core chemistry found toward the IRS3 cluster. Higher angular resolution interferometric observations are needed to determine the origin of the rich molecular line emission arising from this source.
  
For IRS5, our SMA observations reveal that this object also presents a hot-core-like chemistry, indicative of its youth. The molecular abundances measured toward IRS5 indeed range from 2$\times$10$^{-9}$ for CH$_3$CN to 2$\times$10$^{-6}$ for CH$_3$OH, showing that they have been enhanced by several orders of magnitude with respect to the quiescent gas in molecular dark clouds, as expected in hot cores \citep[see e.g.][]{vandertak00,vandertak03}. Supporting this idea, the derived excitation temperature of the gas toward IRS5 is typical of hot cores ($\sim$150$\,$K, see Section$\,$\ref{subsubsec:molecular gas properties}), indicating that the rich chemistry observed toward this source is a consequence of the evaporation of large amounts of molecular material from the icy mantles of dust grains.

Sulfur-bearing molecules are considered good chemical clocks since their abundance are predicted to vary by several orders of magnitude with time \citep[e.g.][]{charnley97,viti04}. Sulfur chemistry in hot cores starts with the injection of H$_2$S into the gas phase after grain mantle evaporation \citep[although H$_2$S is undetected in solid form in the ISM, this species is believed to be the most abundant carrier of atomic sulfur on the surface of dust grains;][]{smith91}. H$_2$S is destroyed by H and H$_3$O$^+$ to form S and H$_3$S$^+$, which then react with O and OH to form SO \citep{charnley97}. For temperatures $<$300$\,$K, SO is destroyed in favour of SO$_2$ through the reaction O + SO $\rightarrow$ SO$_2$ + photon, which takes over for time-scales on the order of 10$^4$$\,$yr, yielding a $\chi$(SO)/$\chi$(SO$_2$) abundance ratio $\leq$1 \citep{wakelam04}. Keeping in mind the uncertainties in our abundance calculations, the abundance ratio $\chi$(SO)/$\chi$(SO$_2$)$\sim$0.4 measured toward IRS5 implies the age of this source to be at least 10$^4$$\,$yr. This chemical timescale is in reasonable agreement with the age estimated from the SED fitting (Section$\,$\ref{subsubsec:12cocompwithIR}). The H$_2$ densities are likely $\geq$10$^6$$\,$cm$^{-3}$, since SO$_2$ is never found to be more abundant than SO for lower H$_2$ densities \citep{wakelam04}. The derived abundances of OCS and CS ($\sim$8$\times$10$^{-8}$ and $\sim$10$^{-8}$, respectively) are also similar to those predicted by chemical models for the same time-scales of $\geq$10$^4$$\,$yr \citep[see][]{charnley97,wakelam04}.

\subsection{Competitive accretion as possible origin of the MonR2 star cluster}

Previous near-IR observations carried out by \citet{hodapp07} toward the MonR2 star cluster reported a population of H$_2$ jets surrounding the MonR2 clump center. From this, \citet{carpenterreview08} proposed that the observed distribution of H$_2$ 2.12$\,$$\mu$m outflows was consistent with a scenario where star formation had been triggered by the interaction of the large-scale CO outflow with the surrounding ambient cloud. Our SMA observations toward this cluster, although limited to the inner 1$'$-region toward MonR2 (or $\sim$0.25$\,$pc at 830$\,$pc), reveal a population of 11 new CO outflows that appear widely distributed across the central region of the clump. This population of CO outflows, which was not detected in near-IR or mid-IR surveys mainly due to the high extinction toward the cluster center \citep[][]{carpenter97,gutermuth05}\footnote{The Spitzer data also suffer from saturation and crowding toward the MonR2 cloud center \citep[see][]{gutermuth05}.}, is likely associated with the youngest population of Class 0 sources in the cluster. Since this population is widespread across the MonR2 clump, an alternative mechanism different from triggering must dominate the star formation process in the MonR2 cluster.  
  
By inspecting the distribution of Class I and Class II sources in MonR2, Gutermuth (2005) found that the Class I sources cluster around the MonR2 region along the north-south direction while the more evolved Class II objects are distributed across the cluster \citep[see Figure$\,$13 in the review paper of][]{carpenterreview08}. In addition, the distribution of molecular gas, which was derived from the multi-transition analysis of CS \citep{choi00} and the column density of CO \citep{xie94}, peaks toward the center of the cluster.
This configuration would be consistent with the competitive accretion scenario where the youngest (and the most deeply embedded) YSOs, probed by the CO outflows detected with the SMA, form at the center of the MonR2 cluster as a consequence of gas accretion by the potential well. This would explain not only the differences in the distribution of Class 0, I and II sources toward MonR2, but also the centrally-peaked gas distribution toward the cluster center.  

If competitive accretion were the dominant mechanism for star formation in the MonR2 cluster, one would expect the most massive objects in the cluster (IRS 1, IRS 2, IRS 3 and IRS 5) to grow via gas accretion accompanied by a significant population of low mass stars \citep[][]{bonnell06}. As seen in Figure~$\,$\ref{fig:12cooutflows}, the density of low-mass stars in MonR2 is found to be higher around the most massive objects in the cluster such as IRS1, IRS2 and IRS3. A detailed study of the distribution of the stellar density in MonR2 (Rivilla et al. 2015, submitted) indeed shows that this distribution is centrally peaked coinciding with the location of the IRS1, IRS2, IRS3 and IRS5 sources. The maximum value of the stellar density distribution ($>$10$^{5.5}$$\,$stars$\,$pc$^{-3}$; Rivilla et al. 2015, submitted) is found close to the location of the IRS2 source, which corresponds approximately to the geometrical center of the cluster. Note that the most massive stars in the IRS1, IRS2 and IRS3 sub-clusters have comparable masses $\sim$12-15~M$_\odot$ \citep{preibisch02, jimenez-serra13, rivilla14}. The presence of a few scattered sub-clusters hosting massive stars is an expected outcome of the competitive accretion scenario, as shown by the simulations of \citet{bonnell11}. 

The competitive accretion scenario could also explain the large-scale CO outflow detected in MonR2, which emanates from the cluster but for which no powering source has been identified yet \citep[][]{tafalla97,giannakopoulou97}. Since the stellar densities at the center of the MonR2 clump are very high, close dynamical encounters are likely to occur as proposed for the large-scale and powerful CO outflows detected in the Orion BN/KL and DR 21 high-mass star forming regions \citep[see][]{zapata09,zapata13,rivilla13a,rivilla14}. A large-scale SMA mosaic covering the extent of the MonR2 CO outflow is needed to verify this hypothesis.

\section{Conclusions}
\label{sec:conclusions}

Our SMA study of the MonR2 clump center has characterized the thermal continuum and molecular line emission from the main sources in the region at (sub-)millimeter wavelengths. The main conclusions of this study are:

\begin{itemize}
\item Continuum observations: IRS2 and IRS5 are the brightest continuum sources detected at 1.3~mm and 0.85~mm towards MonR2 (IRS3 falls outside the primary beam of the observations, preventing us from accurately characterizing its physical properties). While IRS1 and IRS2 are dominated by free-free emission ($\alpha_{\text{IRS1}} \sim -0.5$ and $\alpha_{\text{IRS2}} \sim -0.2$), IRS5 is dominated by dust thermal emission ($\alpha_{\text{IRS5}} \sim 3.0$.) We obtain a mass estimate of $\sim 0.1-0.3$~M$_\Sun$ for IRS5, and a lower limit of $\sim 0.05-0.15$~M$_\Sun$ for IRS3, the most luminous source in the region. Higher-angular resolution observations in the SMA Very Extended configuration reveal the substructure of IRS5 and IRS1 at the 0.5'' level. IRS2 however retains its compact morphology at this resolution.
\item Molecular outflows: We detect 11 previously unknown young outflows in the \super{12}CO (2-1) transition. IRS5 and IRS3 are associated with bright collimated bipolar outflows, with masses in the 10\super{-3}-10\super{-2} M\sub{$\Sun$} range. Their momenta are of order 10\super{-2} - 10\super{-1} M\sub{$\Sun$}~km~s\super{-1}, while the energies are of order $\sim$0.3-2.5 M\sub{$\Sun$}~km~\super{2}~s\super{-2}. A comparison with known IR sources indicates that the IRS5 outflow is likely associated with the IRS5-A source and that the IRS3 outflow is likely driven by the near IR source IRS3B, whose mass was reported to be 8-12 M\sub{$\Sun$} \citep{preibisch02}. The SED fitting for IRS5 yields a stellar mass and luminosity of $\sim 7$~M\sub{$\Sun$} and $\sim 300$~L\sub{$\Sun$}, respectively, and further suggests that IRS5 is a deeply embedded young stellar object still accreting gas.
\item IRS5 chemistry: Our observations show that, besides IRS3, IRS5 is the most chemically active source in the region. From the methyl cyanide $K$ ladder transitions, we derive excitation temperatures of $144 \pm 15$~K for IRS5 and of $126 \pm 22$~K for IRS3. We also provide estimates for the molecular column densities and abundances for the molecular species detected towards IRS5. 
\end{itemize}
We conclude that IRS5 is an intermediate- to high-mass star at an early stage of evolution, as shown by its rich hot-core-like chemistry and the bright, collimated molecular outflow originating from it. 
The CO outflows are found to contribute to the maintenance of turbulence in the MonR2 clump, helping to support it against a global rapid collapse. Together, the newly discovered outflows might have significantly contributed to the velocity dispersion in the MonR2 clump and therefore in the broadening of the molecular linewidths toward the clump center.
 Finally, the detection of 11 new outflows widespread across the MonR2 clump would be consistent with a scenario where star formation in the MonR2 clump takes place through competitive accretion as opposed to triggering.

\acknowledgments
I.J.-S. acknowledges financial support from the People Programme (Marie Curie Actions) of the European Union's Seventh Framework Programme (FP7/2007-2013) under REA grant agreement number PIIF-GA-2011-301538. V.M.R. acknowledges financial support from the Spanish project AYA2010-21697-C05-01.

\newpage
\appendix

\section{Rotational Diagram Method for the CH\sub{3}CN rotational lines}
\label{sec:appendixa}

We detect seven rotational transitions of the methyl cyanide (CH\sub{3}CN) molecule, commonly used to derive the kinetic temperature in star-forming regions, toward MonR2. Images for the CH\sub{3}CN $k=0$, $k=2$ and $k=4$ transitions are presented in Figure \ref{fig:12Linefig}. In Table \ref{tab:mollines} we present the parameters extracted from fitting each $K = 0$ to $K=6$ transition to a Gaussian emission line, as well as upper limits on the integrated and peak intensities for the undetected $K=7$ transition.  The $K=0$ (220.74726~GHz) and $K=1$ (220.74301~GHz) transitions are very close in frequency and hence the fits suffer from partial merging of the lines.

We apply the rotational diagram method to CH\sub{3}CN in order to estimate the gas excitation temperature of IRS5. To this end we assume the emission is optically thin, and that the medium is in Local Thermodynamic Equilibrium (LTE). This implies that the level populations are described by the Maxwell-Boltzmann distribution with temperature $T_{\text{rot}}$, such that:
\begin{equation}
\frac{N_u}{N_l} = \frac{g_u}{g_l} e^{-h\nu/k T_\text{rot}}
\end{equation}
where $g_u$ and $g_l$ designate the statistical weights of the upper and lower levels respectively, and $\nu$ is the transition frequency. Rewriting the expression after assuming LTE gives:
\begin{equation}
\frac{N_u}{g_u} = \frac{N_\text{tot}}{Q(T_\text{rot})} e^{-E_u/k T_\text{rot}}
\label{eq:rotdiageq}
\end{equation}
where $Q(T)$ is the partition function, $E_u$ is the upper level energy, and $N_\text{tot}$ is the total column density. The quantity $\frac{N_u}{g_u}$ is calculated following \citet{zhang98}:
\begin{equation}
\frac{N_{JK}}{g_{JK}} = 1.669\times10^{17}  \frac{\int T_B dv (\text{K km s\super{-1}})}{S_{JK}\mu (\text{D})^2 \nu \text{(MHz)}} \,\, cm^{-2}
\label{eq:Nugu}
\end{equation}
where $\mu$ is the dipole moment and equals 3.92~Debye \citep{blake87}; $S_{JK}$ is the line strength and is given by $S_{JK} = \bra{J^2 - K^2}/J$ ($J = 12$, K = 0 to K = 6 here); and $T_B$ is the brightness temperature. The velocity-integrated brightness temperature can be obtained from the line areas though the Rayleigh-Jeans law:
\begin{equation}
\int T_B dv = 1.224\times10^{6} \quad \frac{S_\nu }{\nu \text{(GHz)$^2$} \theta_B (\arcsec)^2} \,\, K\,km\,s^{-1}
\label{eq:TB}
\end{equation}
where $S_\nu$ is the line area in Jy/beam~km~s\super{-1} as in Table \ref{tab:mollines}, and $\theta_B$ is the geometric mean of the major and minor axis of the beam in arcseconds. The statistical weights are given by $g_{JK} = (2J+1) g_K$ where $g_K = 4$ for K = 0 or K $\neq 3n$, and $g_K = 8$ for K = $3n$, where $n$ is a nonzero integer. 

The resulting rotational diagram is shown in Figure \ref{fig:rotationaldiag}. Extracting the slope from a linear fit to the data following Equation \ref{eq:rotdiageq} gives an estimate of the gas excitation temperature of $144 \pm 15$~K. The error of T$_{ex}$ has been derived by propagating errors. This value is consistent with typical temperatures of hot cores \citep[e.g.][]{jimenez-serra09}. The total column density can easily be estimated from the y-intercept of the fit. The CH\sub{3}CN partition function is given by \citet{blake87}:
\begin{equation}
Q(T_\text{rot}) = \frac{2}{3} \sqrt{ \bra{\frac{k}{h}}^3 \frac{\pi}{ A B^2} } \quad T_\text{rot}^{3/2}
\end{equation}
$A$ and $B$ are the rotational constants, with A = 158099~MHz and B = 9198.9~MHz from the Cologne Database for Molecular Spectroscopy (CDMS; Mueller et al. 2005). This yields an estimate for the column density of $(3.2 \pm 0.4) \times 10^{13}$~cm\super{-2}.


\newpage
\begin{thebibliography}{}
\bibitem[Alvarez et al.~(2004)]{alvarez04} Alvarez, C., Hoare, M., Glindemann, A., \& Richichi, A. 2004, A\&A, 427, 505
\bibitem[Andersen et al.~(2006)]{andersen06} Andersen, M., Meyer, M. R., Oppenheimer, B. Dougados, C., \& Carpenter, J. 2006, AJ, 132, 2296
\bibitem[Arce et al.~(2010)]{arce10} Arce, H. G., Borkin, M. A., Goodman, A. A., Pineda, J. E., \& Halle, M. W. 2010, ApJ, 715, 1170
\bibitem[Aspin \& Walther~(1990)]{aspin90} Aspin, C., \& Walther, D. M. 1990, A\&A, 235, 387
\bibitem[Baraffe \& Chabrier~(2010)]{baraffe10} Baraffe, I., \& Chabrier, G. 2010, A\&A, 521A, 44B 
\bibitem[Blake et al.~(1987)]{blake87} Blake, G. A., Sutton, E. C., Masson, C. R., \& Phillips, T. G. 1987, ApJ, 315, 621
\bibitem[Beckwith et al.~(1976)]{beckwith76} Beckwith, S., Evans II, N. J., Becklin, E. E., \& Neugebauer, G. 1976, ApJ, 208, 390
\bibitem[Bernasconi \& Maeder~(1996)]{bernasconi96} Bernasconi, P. A., \& Maeder, A. 1996, A\&A, 307, 829
\bibitem[Beuther et al.~(2002)]{beuther02} {Beuther}, H., {Schilke}, P., {Sridharan}, T.~K., {Menten}, K.~M., {Walmsley}, C.~M., \& {Wyrowski}, F. 2002, A\&A, 383, 892
\bibitem[Beuther et al.~(2007b)]{beuther07} Beuther, H., Leurini, S., Schilke, P., et al. 2007b, A\&A, 466, 1065
\bibitem[Bonnell \& Bate~(2006)]{bonnell06} Bonnell, I. A., \& Bate, M. R. 2006, MNRAS, 370, 488
\bibitem[Bonnell et al.~(2011)]{bonnell11} Bonnell, I. A., Smith, R., Clark, P. C., \& Bate, M. R. 2011,  MNRAS, 410, 2339
\bibitem[Boogert et al.~(2008)]{boogert08} Boogert, A. C. A., Pontoppidan, K. M., Knez, C., et al. 2008, ApJ, 678, 985 
\bibitem[Boonman et al.~(2003)]{boonman03} Boonman, A. M. S., Doty, S. D., van Dishoeck, E. F., et al. 2003, A\&A, 406, 937
\bibitem[Boonman \& van Dishoeck~(2003)]{boonmanvandishoeck03} Boonman, A. M. S., \& van Dishoeck, E. F. 2003, A\&A, 403, 1003
\bibitem[Carpenter et al.~(1997)]{carpenter97} Carpenter, J. M., Meyer, M. R., Dougados, C., Strom, S. E., \& Hillenbrand, L. A. 1997, AJ, 114, 198
\bibitem[Carpenter \& Hodapp~(2008)]{carpenterreview08} Carpenter, J.~M., \& Hodapp, K.~W.\ 2008, Handbook of Star Forming Regions vol.~1, 899-927 (2008), 899 
\bibitem[Carroll et al.~(2009)]{carroll09} Carroll, J. J., Frank, A., Blackman, E. G., Cunningham, A. J., \& Quillen, A. C. 2009, ApJ, 695, 1376
\bibitem[Charnley et al.~(1992)]{charnley92} Charnley, S. B., Tielens, A. G. G. M., \& Millar, T. J. 1992, ApJ, 399, L71
\bibitem[Charnley~(1997)]{charnley97} Charnley, S. B. 1997, ApJ, 481, 396
\bibitem[Choi et al.~(2000)]{choi00} Choi, M., Evans, II, N. J., Tafalla, M., \& Bachiller, R. 2000, ApJ, 538, 738
\bibitem[Cohen \& Frogel~(1977)]{cohen77} Cohen, J. G., \& Frogel, J. A. 1977, ApJ, 211, 178
\bibitem[de Wit et al.~(2009)]{dewit09} de Wit, W. J., Hoare, M. G., Fujiyoshi, T., et al. 2009, A\&A, 494, 157
\bibitem[Downes et al.~(1975)]{downes75} Downes, D., Winnberg, A., Goss, W. M., \& Johansson, L. E. B. 1975, A\&A, 44, 243
\bibitem[Dunham et al.~(2014)]{dunham14} Dunham, M. M., Arce, H. G., Mardones, D. et al. 2014, ApJ, 783, 29
\bibitem[Enoch et al.~(2006)]{enoch06} Enoch, M. L., Young, K. E., Glenn, J., et al. 2006, ApJ, 638, 293 
\bibitem[Garden et al.~(1991)]{garden91} Garden, R. P., Hayashi, M., Gatley, I., Hasegawa, T., \& Kaifu, N. 1991, ApJ, 374, 540
\bibitem[Genzel \& Downes~(1977)]{genzel77} Genzel, R. \& Downes, D. 1977, A\&A, 61, 117
\bibitem[Giannakopoulou et al.~(1997)]{giannakopoulou97} Giannakopoulou, J., Mitchell, G. F., Hasegawa, T. I., Matthews, H. E., \& Maillard, J.-P. 1997, ApJ, 487, 346
\bibitem[Goldsmith \& Langer~(1999)]{goldsmith99} Goldsmith, P. F., \& Langer, W. D. 1999, ApJ, 517, 209
\bibitem[Gonatas et al.(1992)]{gonatas12} Gonatas, C.~P., Palmer, P., \& Novak, G.\ 1992, \apj, 398, 118 
\bibitem[Gueth \& Guilloteau~(1999)]{gueth99} Gueth, F. \& Guilloteau, S. 1999, A\&A, 343, 571
\bibitem[Guilloteau et al.~(1992)]{guilloteau92} Guilloteau, S., Bachiller, R., Fuente, A., \& Lucas, R. 1992, A\&A, 265, L49
\bibitem[Gutermuth et al.~(2005)]{gutermuth05} Gutermuth, R. A., Megeath, S. T., Pipher, J. L., et al. 2005, ApJ, 632, 397
\bibitem[Gutermuth et al.~(2009)]{gutermuth09} Gutermuth, R. A., Megeath, S. T., Myers, P. C., et al. 2009, ApJS, 184, 18
\bibitem[Gutermuth et al.~(2011)]{gutermuth11} Gutermuth, R. A., Pipher, J. L., Megeath, S. T., et al. 2011, ApJ, 739, 84
\bibitem[Hackwell et al.~(1982)]{hackwell82} Hackwell, J. A., Grasdalen, G. L., \& Gehrz, R. D. 1982, ApJ, 252, 250
\bibitem[Henning et al.~(1992)]{henning92} Henning, Th., Chini, R., \& Pfau W. 1992, A\&A, 263, 285
\bibitem[Herbst \& Racine~(1976)]{herbst76} Herbst, W., \& Racine, R. 1976, AJ, 81, 840 
\bibitem[Hildebrand~(1983)]{hildebrand83} Hildebrand, R. H. 1983, QJRAS, 24, 267
\bibitem[Hirano et al.~(2006)]{hirano06} Hirano, N., Liu, S.-Y., Shang, H., et al. 2006, ApJ, 636, L141	
\bibitem[Hirano et al.~(2010)]{hirano10} Hirano, N., Ho, P. P. T., Liu, S.-Y., et al. 2010, ApJ, 717, 58 
\bibitem[Ho et al.~(2004)]{ho04} Ho, P. T. P., Moran, J. M., \& Lo, K. Y. 2004, ApJ, 616, L1
\bibitem[Hodapp~(2007)]{hodapp07} Hodapp, K. W. 2007, AJ, 134, 2020
\bibitem[Howard et al.~(1994)]{howard94} Howard, E. M., Pipher, J. L., \& Forrest, W. J. 1994, ApJ, 425, 707
\bibitem[Jim\'{e}nez-Serra et al.~(2009)]{jimenez-serra09} Jim\'{e}nez-Serra, I., Mart\'{i}n-P\'{i}ntado, J., Caselli, P., et al. 2009, ApJ, 703, L157
\bibitem[Jim\'{e}nez-Serra et al.~(2013)]{jimenez-serra13} Jim\'{e}nez-Serra, I., B\'{a}ez-Rubio, A., Rivilla, V. M., et al. 2013, ApJ, 764, L4
\bibitem[Kohno et al.~(2002)]{kohno02} Kohno, M., Koyama, K., \& Hamaguchi, K. 2002, ApJ, 567, 423
\bibitem[Kraemer et al.~(2001)]{kraemer01} Kraemer, K. E., Jackson, J. M., Deutsch, L. K., et al. 2001, ApJ, 561, 282
\bibitem[Krumholz et al.~(2009)]{krumholz09} Krumholz, M. R., Klein, R. I., McKee, C. F., Offner, S., \& Cunningham, A. J. 2009, Sci, 323, 754
\bibitem[Li \& Nakamura~(2006)]{li06} Li, Z.-Y., \& Nakamura, F. 2006, ApJ, 640, L187
\bibitem[Lee et al.~(2007)]{lee07} {Lee}, C.-F., {Ho}, P.~T.~P., {Hirano}, N., et al. 2007, ApJ, 659, 499
\bibitem[Lee et al.~(2010)]{lee10} Lee, C.-F., Hasegawa, T. I., Hirano, N., et al. 2010, ApJ, 713, 731
\bibitem[Loren~(1981)]{loren81} Loren, R. B. 1981, ApJ, 249, 550
\bibitem[Mart\'{i}n-P\'{i}ntado et al.~(2005)]{martin-pintado05} Mart\'{i}n-P\'{i}ntado, J., Jim\'{e}nez-Serra, I., Rodr\'{i}guez-Franco, A. \& Thum, C. 2005, ApJ, 628, L61
\bibitem[Massi et al.~(1985)]{massi85} Massi, M., Felli, M., \& Simon, M. 1985, A\&A, 152, 387
\bibitem[McKee \& Tan~(2003)]{mckee03} McKee, C. F., \& Tan, J. C. 2003, ApJ, 585, 850
\bibitem[Mueller et al.~(2002)]{mueller02} Mueller, K. E., Shirley, Y. L., Evans, N. J., II, \& Jacobson, H. R. 2002, ApJS, 143, 469
\bibitem[Mueller et al.~(2005)]{mueller05} Mueller, H. S. P., Schloeder, F., Stutzki, J., \& Winnewisser, G. 2005, 
J. Mol. Struct., 742, 215
\bibitem[Nakajima et al.~(2003)]{nakajima03} Nakajima, H., Imanishi, K., Takagi, S.-I., Koyama, K., \& Tsujimoto, M. 2003, PASJ, 55, 635
\bibitem[Nakamura \& Li~(2007)]{nakamura07} Nakamura, F., \& Li, Z.-Y. 2007, ApJ, 662, 395 
\bibitem[Palau et al.~(2006)]{palau06} Palau, A., Ho, P. T. P., Zhang, Q., et al. 2006, ApJ, 636, L137
\bibitem[Preibisch et al. ~(2002)]{preibisch02} Preibisch, T., Balega, Y. Y., Schertl, D., \& Weigelt, G. 2002, A\&A, 392, 945
\bibitem[Qiu et al.~(2007)]{qiu07} Qiu, K., Zhang, Q., Beuther, H., \& Yang, J. 2007, ApJ, 654, 361
\bibitem[Qiu \& Zhang~(2009)]{qiu09} Qiu, K., \& Zhang, Q. 2009, ApJ, 702, L66
\bibitem[Qiu et al.~(2009)]{qiu09b} {Qiu}, K., {Zhang}, Q.,  {Wu}, J., \& {Chen}, H.-R. 2009, ApJ, 696, 66
\bibitem[Reipurth \& Bally~(2001)]{reipurth01} Reipurth, B. \& Bally, J. 2001, ARA\&A, 39, 403
\bibitem[Ridge et al.~(2003)]{ridge03} Ridge, N. A., Wilson, T. L., Megeath, S. T., Allen, L. E., \& Myers, P. C. 2003, AJ, 126, 286
\bibitem[Rivilla et al.~(2013a)]{rivilla13a} Rivilla, V. M., Mart\'{i}n-P\'{i}ntado, J., Jim\'{e}nez-Serra, I., \& Rodr\'{i}guez-Franco, A. 2013, A\&A, 554, A48
\bibitem[Rivilla et al.~(2013b)]{rivilla13b} Rivilla, V. M., Mart\'{i}n-P\'{i}ntado, J., Sanz-Forcada, J., Jim\'{e}nez-Serra, I., \& Rodr\'{i}guez-Franco, A. 2013, MNRAS, 434, 2313
\bibitem[Rivilla et al.~(2014)]{rivilla14} Rivilla, V. M., Jim\'{e}nez-Serra, I., \& Mart\'{i}n-P\'{i}ntado, J. \& Sanz-Forcada, J. 2014, MNRAS, 437, 1561
\bibitem[Roberts et al.~(2010)]{roberts10}	Roberts, J. F., Jim\'enez-Serra, I., Mart\'{i}n-Pintado, J., et al. 2010, A\&A, 513A, 64R
\bibitem[Robitaille et al.~(2006)]{robitaille06} Robitaille, T. P., Whitney, B. A., Indebetouw, R., Wood, K., \& Denzmore, P. 2006, ApJS, 167, 256
\bibitem[Scoville et al.~(1986)]{scoville86} Scoville, N. Z., Sargent, A. I., Sanders, D. B., et al. 1986, ApJ, 303, 416
\bibitem[Siess et al.~(2000)]{siess00} Siess, L., Dufour, E., \& Forestini, M. 2000, A\&A, 358, 593
\bibitem[Smith~(1991)]{smith91} Smith, R. G. 1991, MNRAS, 249, 172
\bibitem[Smits et al.~(1998)]{smits98} Smits, D. P., Cohen, R. J., \& Hutawarakorn, B. 1998, MNRAS, 296, L11
\bibitem[Sridharan et al.~(2002)]{sridharan02} Sridharan, T. K., Beuther, H., Schilke, P., Menten, K. M., \& Wyrowski, F. 2002, ApJ, 566, 931
\bibitem[Stanke \& Williams~(2007)]{stanke07} Stanke, T., \& Williams, J. P. 2007, AJ, 133, 1307
\bibitem[Tafalla et al.~(1997)]{tafalla97} Tafalla, M., Bachiller, R., Wright, M. C. H., \& Welch, W.J., 1997, ApJ, 474, 329
\bibitem[Thronson et al.~(1980)]{thronson80} Thronson, H. A. Jr., Gatley, I., Harvey, P. M., Sellgren, K., \& Werner, M. W. 1980, ApJ, 237, 66
\bibitem[Torrelles et al.~(1983)]{torrelles83} Torrelles, J. M., Rodr\'{i}guez, L. F., Canto, J., et al. 1983, ApJ, 274, 214
\bibitem[van der Tak et al.~(2000)]{vandertak00} van der Tak, F. F. S., van Dishoeck, E. F., \& Caselli, P. 2000, A\&A, 361, 327
\bibitem[van der Tak et al.~(2003)]{vandertak03} van der Tak, F. F. S., Boonman, A. M. S., Braakman, R., \& van Dishoeck, E. F. 2003, A\&A, 412, 133
\bibitem[Viti et al.~(2004)]{viti04} Viti, S., Collings, M. P., Dever, J. W., McCoustra, M. R. S., \& Williams, D. A. 2004, MNRAS, 354, 1141
\bibitem[Wakelam et al.~(2004)]{wakelam04} Wakelam, V., Caselli, P., Ceccarelli, C., Herbst, E., \& Castets, A.  2004, A\&A, 422, 159
\bibitem[Walker et al.~(1990)]{walker90} Walker, C. K., Adams, F. C., \& Lada, C. J. 1990, ApJ, 349, 515
\bibitem[Wang et al.~(2011)]{wang11} Wang, K., Zhang, Q., Wu, Y., \& Zhang, H. 2011, AJ, 735, 64
\bibitem[Wilson~(1999)]{wilson99} Wilson, T. L. 1999, Rep. Prog. Phys. 62 143
\bibitem[Wolf et al.~(1990)]{wolf90} Wolf, G. A., Lada, C. J., \& Bally, J. 1990, AJ, 100, 1892
\bibitem[Wood \& Churchwell~(1989)]{wood89} Wood, D. O. S., \& Churchwell, E. 1989, ApJS, 69, 831
\bibitem[Xie \& Goldsmith~(1994)]{xie94} Xie, T., \& Goldsmith, P. F. 1994, ApJ, 430, 252
\bibitem[Yorke \& Sonnhalter~(2002)]{yorke02} Yorke, H. W., \& Sonnhalter, C. 2002, ApJ, 569, 846
\bibitem[Zapata et al.~(2009)]{zapata09} Zapata, L. A., Menten, K., Reid, M., \& Beuther, H. 2009, ApJ, 691, 332
\bibitem[Zapata et al.~(2013)]{zapata13} Zapata, L. A., Schmid-Burgk, J., P\'{e}rez-Goytia, N., et al. 2013, ApJ, 765, L29
\bibitem[Zhang et al.~(1998)]{zhang98} Zhang, Q., Ho, P. T. P., \& Ohashi, N. 1998, ApJ, 494, 636
\bibitem[Zhang et al.~(2000)]{zhang00} Zhang, Q., Ho, P. T. P., \& Wright, M.~C.~H. 2000, AJ, 119, 1345
\bibitem[Zhang et al.~(2001)]{zhang01} {Zhang}, Q., {Hunter}, T.~R., {Brand}, J., et al. 2001, ApJ, 552, L167
\bibitem[Zhang et al.~(2005)]{zhang05} {Zhang}, Q., {Hunter}, T.~R., {Brand}, J., et al. 2005, ApJ, 625, 864
\bibitem[Zhu et al.~(2010)]{zhu10} Zhu, Z., Hartmann, L., Gammie, C. F., et al. 2010, ApJ, 713, 1134
\end{thebibliography}
\end{document}